\documentclass[fleqn,usenatbib]{mnras}

\usepackage{newtxtext,newtxmath}
\usepackage{longtable}
\usepackage[T1]{fontenc}

\DeclareRobustCommand{\VAN}[3]{#2}
\let\VANthebibliography\thebibliography
\def\thebibliography{\DeclareRobustCommand{\VAN}[3]{##3}\VANthebibliography}

\usepackage{graphicx}	
\usepackage{amsmath}	
\usepackage{caption}
\usepackage{subcaption}
\usepackage{lineno}

\newcommand{\Lagr}{\mathcal{L}}

\title[DESI Peculiar Velocity Survey -- Fundamental Plane]{DESI Peculiar Velocity Survey -- Fundamental Plane}

\author[K. Said et al.]{Khaled Said,$^{1}$\thanks{E-mail: k.saidahmedsoliman@uq.edu.au}
Cullan Howlett,$^{1}$
Tamara Davis,$^{1}$
John Lucey,$^{2}$
Christoph Saulder,$^{3}$
Kelly Douglass,$^{4}$
\newauthor
Alex G. Kim,$^{5}$
Anthony Kremin,$^{5}$
Caitlin Ross,$^{1}$
Greg Aldering,$^{5}$
Jessica Nicole Aguilar,$^{5}$
Steven Ahlen,$^{6}$
\newauthor
Segev BenZvi,$^{4}$
Davide Bianchi,$^{7}$
David Brooks,$^{8}$
Todd Claybaugh,$^{5}$
Kyle Dawson,$^{9}$
Axel de la Macorra,$^{10}$
\newauthor
Biprateep Dey,$^{11}$
Peter Doel,$^{8}$
Kevin Fanning,$^{13,12}$
Simone Ferraro,$^{5,14}$
Andreu Font-Ribera,$^{8,15}$
\newauthor
Jaime E. Forero-Romero,$^{17,16}$
Enrique Gaztañaga,$^{20,19,18}$
Satya Gontcho A Gontcho,$^{5}$
Julien Guy,$^{5}$
\newauthor
Klaus Honscheid,$^{21,22,23}$
Robert Kehoe,$^{24}$
Theodore Kisner,$^{5}$
Andrew Lambert,$^{5}$
Martin Landriau,$^{5}$
\newauthor
Laurent Le Guillou,$^{25}$
Marc Manera,$^{26,15}$
Aaron Meisner,$^{27}$
Ramon Miquel,$^{28,15}$
John Moustakas,$^{29}$
\newauthor
Andrea Muñoz-Gutiérrez,$^{10}$
Adam Myers,$^{30}$
Jundan Nie,$^{31}$
Nathalie Palanque-Delabrouille,$^{32,5}$
\newauthor
Will Percival,$^{33,34,35}$
Francisco Prada,$^{36}$
Graziano Rossi,$^{37}$
Eusebio Sanchez,$^{38}$
David Schlegel,$^{5}$
\newauthor
Michael Schubnell,$^{39,40}$
Joseph Harry Silber,$^{5}$
David Sprayberry,$^{27}$
Gregory Tarlé,$^{40}$
Mariana Vargas Magana,$^{10}$
\newauthor
Benjamin Alan Weaver,$^{27}$
Risa Wechsler,$^{41,13,12}$
Zhimin Zhou,$^{31}$
Hu Zou,$^{31}$
\\
$^{1}$School of Mathematics and Physics, University of Queensland, 4072, Australia\\
$^{2}$Institute for Computational Cosmology, Department of Physics, Durham University, South Road, Durham DH1 3LE, UK\\
$^{3}$Max Planck Institute for Extraterrestrial Physics, Gie\ss enbachstra\ss e 1, 85748 Garching, Germany\\
$^{4}$Department of Physics \& Astronomy, University of Rochester, 206 Bausch and Lomb Hall, P.O. Box 270171, Rochester, NY 14627-0171, USA\\
$^{5}$Lawrence Berkeley National Laboratory, 1 Cyclotron Road, Berkeley, CA 94720, USA\\
$^{6}$Physics Dept., Boston University, 590 Commonwealth Avenue, Boston, MA 02215, USA\\
$^{7}$Dipartimento di Fisica ``Aldo Pontremoli'', Universit\`a degli Studi di Milano, Via Celoria 16, I-20133 Milano, Italy\\
$^{8}$Department of Physics \& Astronomy, University College London, Gower Street, London, WC1E 6BT, UK\\
$^{9}$Department of Physics and Astronomy, The University of Utah, 115 South 1400 East, Salt Lake City, UT 84112, USA\\
$^{10}$Instituto de F\'{\i}sica, Universidad Nacional Aut\'{o}noma de M\'{e}xico,  Cd. de M\'{e}xico  C.P. 04510,  M\'{e}xico\\
$^{11}$Department of Physics \& Astronomy and Pittsburgh Particle Physics, Astrophysics, and Cosmology Center (PITT PACC), University of Pittsburgh,\\ 3941 O'Hara Street, Pittsburgh, PA 15260, USA\\
$^{12}$Kavli Institute for Particle Astrophysics and Cosmology, Stanford University, Menlo Park, CA 94305, USA\\
$^{13}$SLAC National Accelerator Laboratory, Menlo Park, CA 94305, USA\\
$^{14}$University of California, Berkeley, 110 Sproul Hall \#5800 Berkeley, CA 94720, USA\\
$^{15}$Institut de F\'{i}sica d’Altes Energies (IFAE), The Barcelona Institute of Science and Technology, Campus UAB, 08193 Bellaterra Barcelona, Spain\\
$^{16}$Departamento de F\'isica, Universidad de los Andes, Cra. 1 No. 18A-10, Edificio Ip, CP 111711, Bogot\'a, Colombia\\
$^{17}$Observatorio Astron\'omico, Universidad de los Andes, Cra. 1 No. 18A-10, Edificio H, CP 111711 Bogot\'a, Colombia\\
$^{18}$Institut d'Estudis Espacials de Catalunya (IEEC), 08034 Barcelona, Spain\\
$^{19}$Institute of Cosmology and Gravitation, University of Portsmouth, Dennis Sciama Building, Portsmouth, PO1 3FX, UK\\
$^{20}$Institute of Space Sciences, ICE-CSIC, Campus UAB, Carrer de Can Magrans s/n, 08913 Bellaterra, Barcelona, Spain\\
$^{21}$Center for Cosmology and AstroParticle Physics, The Ohio State University, 191 West Woodruff Avenue, Columbus, OH 43210, USA\\
$^{22}$Department of Physics, The Ohio State University, 191 West Woodruff Avenue, Columbus, OH 43210, USA\\
$^{23}$The Ohio State University, Columbus, 43210 OH, USA\\
$^{24}$Department of Physics, Southern Methodist University, 3215 Daniel Avenue, Dallas, TX 75275, USA\\
$^{25}$Sorbonne Universit\'{e}, CNRS/IN2P3, Laboratoire de Physique Nucl\'{e}aire et de Hautes Energies (LPNHE), FR-75005 Paris, France\\
$^{26}$Departament de F\'{i}sica, Serra H\'{u}nter, Universitat Aut\`{o}noma de Barcelona, 08193 Bellaterra (Barcelona), Spain\\
$^{27}$NSF NOIRLab, 950 N. Cherry Ave., Tucson, AZ 85719, USA\\
$^{28}$Instituci\'{o} Catalana de Recerca i Estudis Avan\c{c}ats, Passeig de Llu\'{\i}s Companys, 23, 08010 Barcelona, Spain\\
$^{29}$Department of Physics and Astronomy, Siena College, 515 Loudon Road, Loudonville, NY 12211, USA\\
$^{30}$Department of Physics \& Astronomy, University  of Wyoming, 1000 E. University, Dept.~3905, Laramie, WY 82071, USA\\
$^{31}$National Astronomical Observatories, Chinese Academy of Sciences, A20 Datun Rd., Chaoyang District, Beijing, 100012, P.R. China\\
$^{32}$IRFU, CEA, Universit\'{e} Paris-Saclay, F-91191 Gif-sur-Yvette, France\\
$^{33}$Department of Physics and Astronomy, University of Waterloo, 200 University Ave W, Waterloo, ON N2L 3G1, Canada\\
$^{34}$Perimeter Institute for Theoretical Physics, 31 Caroline St. North, Waterloo, ON N2L 2Y5, Canada\\
$^{35}$Waterloo Centre for Astrophysics, University of Waterloo, 200 University Ave W, Waterloo, ON N2L 3G1, Canada\\
$^{36}$Instituto de Astrof\'{i}sica de Andaluc\'{i}a (CSIC), Glorieta de la Astronom\'{i}a, s/n, E-18008 Granada, Spain\\
$^{37}$Department of Physics and Astronomy, Sejong University, Seoul, 143-747, Korea\\
$^{38}$CIEMAT, Avenida Complutense 40, E-28040 Madrid, Spain\\
$^{39}$Department of Physics, University of Michigan, Ann Arbor, MI 48109, USA\\
$^{40}$University of Michigan, Ann Arbor, MI 48109, USA\\
$^{41}$Physics Department, Stanford University, Stanford, CA 93405, USA\\
}

\date{Accepted XXX. Received YYY; in original form ZZZ}

\pubyear{2024}

\begin{document}
\label{firstpage}
\pagerange{\pageref{firstpage}--\pageref{lastpage}}
\maketitle

\begin{abstract}
The Dark Energy Spectroscopic Instrument (DESI) Peculiar Velocity Survey aims to measure the peculiar velocities of early and late type galaxies within the DESI footprint using both the Fundamental Plane and optical Tully-Fisher relations. Direct measurements of peculiar velocities can significantly improve constraints on the growth rate of structure, reducing uncertainty by a factor of approximately 2.5 at redshift 0.1 compared to the DESI Bright Galaxy Survey's redshift space distortion measurements alone. We assess the quality of stellar velocity dispersion measurements from DESI spectroscopic data. These measurements, along with photometric data from the Legacy Survey, establish the Fundamental Plane relation and determine distances and peculiar velocities of early-type galaxies. During Survey Validation, we obtain spectra for 6698 unique early-type galaxies, up to a photometric redshift of 0.15. 64\% of observed galaxies (4267) have relative velocity dispersion errors below 10\%. This percentage increases to 75\% if we restrict our sample to galaxies with spectroscopic redshifts below 0.1. We use the measured central velocity dispersion, along with photometry from the DESI Legacy Imaging Surveys, to fit the Fundamental Plane parameters using a 3D Gaussian maximum likelihood algorithm that accounts for measurement uncertainties and selection cuts. In addition, we conduct zero-point calibration using the absolute distance measurements to the Coma cluster, leading to a value of the Hubble constant, $H_0 = 76.05 \pm 0.35$(statistical) $\pm 0.49$(systematic FP) $\pm 4.86$(statistical due to calibration) $\mathrm{km \ s^{-1} Mpc^{-1}}$. This $H_0$ value is within $2\sigma$ of Planck Cosmic Microwave Background results and within $1\sigma$, of other low redshift distance indicator-based measurements. 
\end{abstract}




\section{Introduction}

The Lambda Cold Dark Matter ($\Lambda$CDM) model stands as the prevailing cosmological framework to describe the Universe. It combines Einstein's cosmological constant, $\Lambda$, representing dark energy \citep{Carroll2001,Peebles2003}, with nonbaryonic cold dark matter (CDM; \citealt{Hut1977,Lee1977,Sato1977,Dicus1977,Vysotski1977}). Despite its consistency with a wide range of cosmological observations, the nature of its two main components, dark matter and dark energy, remains unknown to us \citep{Peebles2021}. Moreover, tensions persist between measurements of the present-day expansion rate of the universe from the early-universe (e.g., \citealt{Planck2020}) and those derived from the late-universe (e.g., \citealt{Riess2022}). Additional tensions arise from measurements of the strength of clustering of matter in the universe between early- and late-universe observations \citep{Hildebrandt2020}. Furthermore, discrepancies emerge from measurements of the bulk flow, directly derived from late-universe observations, which do not align with the expected values from the $\Lambda$CDM model \citep{Courtois2023,Whitford2023,Watkins2023,Hoffman2024}. In addition, current cosmological observations seem to favor a slightly higher growth index, $\gamma$ than the value predicted by general relativity and the $\Lambda$CDM model, which might indicate weaker gravity or suppression of the growth of structure \citep{Nguyen2023}.

Addressing these tensions requires substantial efforts in acquiring additional data to precisely understand the underlying discrepancies. Nevertheless, examining the tension using independent techniques should be our utmost priority, as it involves overcoming potential  sources of systematics that may be linked to new physics \citep{Lahav2021}.

The Dark Energy Spectroscopic Instrument (DESI; \citealt{DESI_Collaboration2022}) is a ground-based spectroscopic survey that is targeting 40 million galaxies and quasars over a 14,000 square degrees footprint. It is  over two years into its 5-year observing campaign. DESI's aims align very closely with the above suggestion of unveiling the nature of dark energy by conducting the most precise measurement of the universe expansion history using the baryon acoustic oscillation (BAO) \citep{Levi2013}. DESI will also measure the growth rate of cosmic structure using redshift space distortions (RSD) \citep{DESI_Collaboration2016}.

The DESI peculiar velocity survey serves as a secondary target program, designed to complement and enhance the primary goal of the DESI survey \citep{Saulder2023}. Peculiar velocities of galaxies can be measured through two primary methods: using distance indicators and velocity field reconstruction based on local density measurements. In this paper, we employ the direct method, specifically aiming to enrich the dataset by incorporating $\sim180,000$ directly measured distances using redshift-independent distance indicators. These directly measured distances provide valuable independent information about galaxy velocities, helping to disentangle the effects of the smooth Hubble flow due to the cosmic expansion and the peculiar velocity due to gravitational interactions on galaxy motion. The peculiar velocity survey offers an independent and complementary approach to the main DESI surveys at higher redshift, allowing for a better understanding of large-scale structure through measurement of $f\sigma_8$ below $z=0.1$ and the overall expansion of the universe through measurements of $H_0$.

Historically, large peculiar velocity surveys, from the Two Micron All Sky Survey (2MTF: \citealt{Masters2008}) and the 6dF Galaxy Survey (6dFGS: \citealt{Springob2014}) have relied on either Tully-Fisher relation \citep{Tully1977} or the Fundamental Plane (FP) relation \citep{Djorgovski1987,Dressler1987} to obtain distance measurements. The DESI peculiar velocity survey will be able to provide observations for these two distance indicators in an unprecedented and innovative way, thanks to the DESI's substantial increase in the survey speed due to the large field of view and the densely populated focal plane with 5000 robotic fiber positioners \citep{DESICollaboration2016b}. 

The FP relation utilizes two distance-independent observables of elliptical galaxies, the mean surface brightness ($I_e$) and the central velocity dispersion ($\sigma_0$) to infer the physical effective radius ($R_e$), which is distance dependent. Comparing the physical radius to the angular effective radius ($\theta_e$) allows for the measurement of distances. Similarly, the Tully-Fisher relation uses the rotational velocity of galaxies as a distance-independent observable to predict the absolute magnitude. Then, by comparing the predicted absolute magnitude with the observed apparent magnitude, one can derive distance estimates for these galaxies. These directly measured distances and peculiar velocities have been instrumental in constraining crucial cosmological parameters.

The 6dF Galaxy Survey's \citep{Jones2009} primary objective was to measure peculiar velocities for a sample of approximately 9,000 early-type galaxies up to redshift $z < 0.055$ using the FP relation, aiming to constrain the growth rate of cosmic structure (6dFGSv: \citealt{Magoulas2012}). \cite{Qin2019} utilized the 6dFGSv data to estimate the Density-Momentum power spectrum and derived a value of $f\sigma_8 = 0.404 \pm 0.082$. Similarly, \cite{Adams2020} employed the 6dFGSv data and applied the cross-covariance between galaxy redshift-space distortions and peculiar velocities to constrain the growth rate of structure, obtaining $f\sigma_8=0.384\pm0.052 \text{(statistical)} \pm 0.061 \text{(systematic)}$. \cite{Said2020} compared a sub-sample of the 6dFGSv FP galaxies along with a sample of SDSS galaxies to the velocity field reconstruction and reported a value of $f\sigma_8 = 0.338 \pm 0.027$. More recently, \cite{Turner2023} used the full 6dFGSv sample to measure galaxy-galaxy, galaxy-velocity, and velocity-velocity auto- and cross-correlation functions, finding a value of $f\sigma_8 = 0.358 \pm 0.075$. While all these measurements, using different methods, are consistent, they all favour a lower value than the predicted value from the Planck $\Lambda$CDM model \citep{Planck2020}.

More recently, Cosmicflows-4 has undertaken the ambitious task of combining almost all previous peculiar velocity surveys into a comprehensive dataset \citep{Tully2023}. This compilation incorporates distances and peculiar velocities for 55,877 galaxies, utilizing eight different methodologies. The largest number of galaxies were derived from two new datasets, with approximately 35,000 galaxies measured using the FP relation \citep{Howlett2022} and about 10,000 galaxies using the Tully-Fisher relation \citep{Kourkchi2020}. The integration of these diverse datasets has resulted in a Hubble constant measurement of $H_0 = 74.6 \pm 0.8$ \text{(statistical)} $\pm 3.0$ \text{(systematic)} \citep{Tully2023}. The Cosmicflows-4 catalogue has proven invaluable for various cosmological measurements, including assessing the impact of our local environment on $H_0$ \citep{Giani2024}, measuring the growth rate of structure \citep{Courtois2023,Boubel2024a}, and developing an improved method for determining the Hubble constant using the Tully-Fisher relation \citep{Boubel2024b,Scolnic2024}. While the statistical errors from these studies are relatively small, thanks to the considerable increase in the sample size, there remain concerns about potential large systematic errors due to the combination of different methods and calibrators. 

This reinforces the need for larger and more homogeneous surveys, like the DESI peculiar velocity survey, to further advance our understanding of cosmological parameters and address potential sources of systematic effects.

The DESI peculiar velocity survey is expected to yield a remarkable number of directly measured distances and peculiar velocities. Specifically, it is projected to obtain 186,000 such measurements, with 133,000 of them using the FP relation and 53,000 using the Tully-Fisher relation\footnote{The rotational velocities required for the Tully-Fisher relation in DEAI-PV are measured using optical spectra from multiple fibers positioned along the galaxy's major axis, distinguishing this approach from the Radio Tully-Fisher approach which uses HI 21 cm line width measurements.} \citep{Saulder2023}. These numbers represent a significant increase in the scale, being about four times all previous peculiar velocity surveys combined \citep{Tully2023}. 

With this extensive dataset, DESI will achieve a new level of precision in constraining cosmological parameters. The large number of directly measured distances and peculiar velocities will enable more robust and statistically significant results compared to all previous surveys. 

This work, accompanied by a parallel work conducted by Douglass et al. in prep, introduces the initial outcomes of the DESI peculiar velocity survey using Survey Validation data. In Douglass et al. in prep, the emphasis is on the Tully-Fisher relation, while this work focuses only on the FP relation. 

During the Survey Validation (SV) phase, our approach involved conducting observations for a randomized subset of our designated targets. This selection addressed key questions: First, we evaluated the achievability of the Signal-to-Noise Ratio (SNR) needed for accurate velocity dispersion measurements across our magnitude and redshift range. Additionally, we examined potential sources of systematic errors in velocity dispersion measurements that might impact the precision of our distance and peculiar velocity determinations. Furthermore, we evaluated the efficiency of our photometric selection in accurately pinpointing genuine elliptical galaxies. 

While the primary objective of this paper is to offer insights into these fundamental questions, we also conducted the fitting of the FP relation. By doing so, we produced the initial catalogue of distances and peculiar velocities from DESI. Moreover, we present our preliminary findings concerning the measurement of the Hubble constant.

This paper is organized as follows: Section \ref{S_intro_FP} provides a description of the FP relation. In Section \ref{S_Data}, we offer an introduction to the dataset employed. Our sample selection process, encompassing photometric and spectroscopic aspects, is explained in Section \ref{S_sample_selection}. The derivation of FP parameters, inclusive of internal and external consistency checks, is presented in Section \ref{S_FP_parameters}. We show the process of fitting the FP relation in Section \ref{S_FP_fit}. In Section \ref{S_zero_point}, we discuss the zero-point calibration process and present absolute distance measurements. Section \ref{S_hubble_constant}, unfolds our measurement of the Hubble constant. Summary and Conclusions are in Sections \ref{S_Discussion} and \ref{S_conclusions}, respectively.

Unless otherwise stated, in this paper we assume a flat $\Lambda$CDM cosmological model with $\Omega_{m}=0.31$ and $H_0=100h$\,km\,s$^{-1}$\,Mpc$^{-1}$. All magnitudes are on the AB magnitude system. All uses of `$\log$' should be taken to mean logarithms taken to the base 10.


\section{Fundamental Plane}
\label{S_intro_FP}
Distance indicators operate on the premise of linking distance-independent parameters such as kinematics, with distance-dependent characteristics like luminosity or size. This relationship allows for the estimation of galaxy distances based on known distance-independent parameters. The first distance-indicator relation for elliptical galaxies was the luminosity-stellar velocity dispersion correlation, known as the Faber-Jackson relation (FJ: \citealt{Faber1976}). It is worth noting that the foundation of this relation was first suggested by \cite{Minkowski1962}, although they regarded it as inadequate probably due to the inclusion of flattened galaxies with high rotational velocities. The FJ relation follows the form $L \propto \sigma^4$. Simultaneously, \cite{Kormendy1977} identified a correlation between surface brightness and size of elliptical galaxies. Both FJ and Kormendy relations did not seem very promising as distance indicators due to their large scatter.

A significant advancement emerged approximately a decade later when it became evident that the FJ and Kormendy relations were special instances of a more general relation known as the FP \citep{Djorgovski1987,Dressler1987}. These works show that a galaxy's effective radius ($R_e$), surface brightness ($I_e$), and stellar velocity dispersion are related through a power-law relationship, expressed as $R_e \propto \sigma^a I_e^b$. 

The obvious explanation of the FP rested upon the virial equilibrium, linking a galaxy's mass ($M$) to its velocity dispersion $(\sigma)$ and effective radius $(R_e)$ through the equation $M \propto \sigma^2 R_e$ \citep{Faber1987}. However, the FP coefficients showed significant deviations from the predictions of the virial theorem \citep{Hudson1997,Colless2001,Bernardi2003,Magoulas2012,Said2020,D'Eugenio2021,Howlett2022}. Varied contributors to this divergence were identified, including the fluctuation of the mass-to-light ratio \citep{Faber1987}, variations in the surface brightness profiles of early-type galaxies \citep{Ciotti1996}, and the proportion of dark matter within the kinematic observation region \citep{Moster2010}. Despite the deviations, the FP's applicability persisted across all early-type galaxies, maintaining a scatter of approximately 0.1 dex, which translates to an accuracy in distance estimation of around 23\% \citep{Lynden-Bell1988}.

While the current work concentrates solely on distance and peculiar velocity measurements, the forthcoming data from DESI holds the promise of delving into sources of the FP's tilt (its deviation from the virial theorem). This potential for detailed analysis arises from the vast and comprehensive datasets that DESI will provide.

\section{Data}
\label{S_Data}
Constructing the FP relation involves utilizing two datasets: photometry and spectroscopy. The photometric data for this work are drawn from the DESI Legacy Imaging Surveys \citep{Dey2019}. The spectroscopic data for our analysis are from the DESI Survey Validation data \citep{DESI_Collaboration2023b}.

The DESI Legacy Imaging Surveys encompass the combination of three individual surveys: the Dark Energy Camera Legacy Survey (DECaLS), the Beijing-Arizona Sky Survey (BASS), and the Mayall z-band Legacy Survey (MzLS). This combination is designed to capture imagery across the expansive 14,000 deg$^2$ footprint ($\delta > -20^\circ$ and $|b| > 15^\circ$) of the DESI survey in three optical bands ($g, r,$ and $z$).

DECaLS, which began observations in August 2014, uses the Dark Energy Camera (DECam; \citealt{Flaugher2015}) at the 4m Blanco telescope at the Cerro Tololo Inter-American Observatory. It covers approximately 9,350 deg$^2$, including 3,580 deg$^2$ in the SGC and 5,770 deg$^2$ in the NGC, complementing the Dark Energy Survey's \citep{DES2005} coverage of 1,130 deg$^2$ within the DESI footprint.

BASS \citep{Zou2017}, which started in spring 2015, images the $\delta > +32^\circ$ region of the DESI NGC footprint (approximately 5,100 deg$^2$) in the $g$ and $r$ optical bands. It utilizes the 90Prime camera \citep{Williams2004}  at the prime focus of the University of Arizona's Bok 2.3m telescope on Kitt Peak.

MzLS complements BASS by imaging the same $\delta > +32\circ$ region of the NGC footprint in the z-band, covering approximately 5,100 deg$^2$.

In addition to optical bands, the Legacy Surveys incorporate mid-infrared photometry from the Wide-field Infrared Survey Explorer (WISE) satellite. WISE conducted an all-sky survey in four bands centered at 3.4, 4.6, 12, and 22 $\mu$m (known as W1, W2, W3, and W4; \citealt{Wright2010}) during its mission.

These surveys collectively contribute to the creation of a photometric catalogue covering the three optical bands ($g, r,$ and $z$) and incorporating the four WISE channels. For the purposes of this paper, our analysis is based on Data Release 9 (DR9) of the Legacy Surveys.

The DESI Early Data Release (EDR) marks the initial public release of DESI spectroscopic data \citep{DESI_Collaboration2023}. It encompasses data from the Survey Validation (SV) phase, conducted between December 2020 and May 2021, prior to the start of the DESI Main Survey. A detailed description of the DESI pipeline can be found in \cite{Guy2023}. For discussion of the survey operations, see \cite{Schlafly2023}. The SV phase consisted of three stages:

\begin{enumerate}
    \item Target Selection Validation (SV1): This phase refined and validated the selection of targets for the Milky Way Survey (MWS), Bright Galaxy Survey (BGS), Luminous Red Galaxies (LRG), Emission Line Galaxies (ELG), and Quasar (QSO) samples. It employed looser target selection cuts and higher signal-to-noise ratios than the Main Survey to optimize selection criteria and survey requirements.
    \item Operations Development (SV2): A brief phase for operational refinements.
    \item 1\% Survey (SV3): This final SV stage further optimized observing procedures and produced high completeness samples over approximately 1\% of the final DESI Main Survey area.
\end{enumerate}

The EDR demonstrates DESI's capabilities, and its advanced instrumentation at the 4m Mayall telescope at Kitt Peak National Observatory. The instrument's wide-field prime focus corrector enables a field of view just over 8 deg$^2$, allowing for efficient large-scale observations. At the heart of DESI's design are 5,020 robotically-controlled fiber positioners \citep{DESICollaboration2016b,Silber2023,Miller2023}, each directing light from individual targets. These fibers feed into ten spectrographs, each equipped with three cameras covering distinct wavelength ranges: B (3600–5800 \AA), R (5760–7620 \AA), and Z (7520–9824 \AA). The spectrographs provide a resolving power that increases from approximately 2000 at 3600 \AA\ to 5500 at 9800 \AA\ 
 \citep{DESI_Collaboration2023}.

In addition to the primary target classes (MWS, BGS, LRG, ELG, and QSO), the EDR also includes observations from DESI's secondary programs, which serve as filler targets.

\section{Sample selection}
\label{S_sample_selection}
As part of the DESI secondary target programs, we provided an initial set of targets to complement the main survey targets. As a result, our sample selection is made in two steps. The initial phase, which we refer to as the photometric selection, was exclusively reliant on the DESI Legacy Imaging Surveys DR9 \citep{Dey2019}, supplemented by photometric redshift catalogues from \cite{Zhou2021}. Subsequently, the second step or the spectroscopic selection involves the integration of spectroscopic data from the DESI survey \citep{DESI_Collaboration2023}. Notably, most of our FP targets are also part of the DESI Bright Galaxy Survey (BGS: \citealt{Ruiz-Macias2020,Hahn2023}). 

A comprehensive description of the photometric sample selection process can be found in \cite{Saulder2023}. For full details of our photometric selection, we refer the reader to that paper. Here, we only provide a brief summary of the photometric selection criteria. Additionally, we will describe how we conducted the spectroscopic sample selection, which facilitated the construction of the peculiar velocity sample for our FP analysis.

\subsection{Photometric Selection}

Our adopted selection criteria are outlined as follows: (1) an $r-$band magnitude cut of $r<18$, which has been corrected for external galactic extinction, ensuring a robust signal-to-noise ratio that subsequently enhances the success rate for velocity dispersion measurements; (2) three distinct colour cuts\footnote{All magnitudes employed in these colour cuts have been adjusted for the external galactic dust extinction.} of 
\begin{eqnarray}
    g-r &>& 0.68\\
    g-r &>& 1.3(r-z) - 0.05\\
    g-r &<& 2(r-z)-0.15.
\end{eqnarray}
These colour cuts serve to eliminate galaxies located beneath the red sequence, dusty galaxies, and peculiar objects like galaxy mergers or galaxies displaying image artefacts, respectively; (3) a circularized radius, which is the half-light radius of the galaxy circularized using the axial ratio, condition of $R_{\text{circ}} > 0$; (4) an axial ratio requirement of $b/a \geq 0.3$; (5) the application of either a de Vaucouleurs or S\'{e}rsic fit to the surface brightness profile, incorporating a S\'{e}rsic index of $n_s > 2.5$, measured collectively across $r$, $g$, and $z$ bands.

These specific criteria for ETGs selection are established based on a combination of past experience in ETG identification from previous surveys like 6dFGSv and SDSS peculiar velocity surveys \citep{Saulder2013,Said2020,Howlett2022}, along with testing involving visual identifications sourced from the Siena Galaxy Atlas (SGA; \citealt{Moustakas2023}) and GalaxyZoo \citep{Lintott2011}

Our photometric selection process, yielding a pool of over 400,000 galaxies deemed suitable for FP analysis. Since our selected sample is from the DESI Legacy Imaging Surveys DR9, it encompasses galaxies extending beyond the bounds of the DESI spectroscopic survey footprint. Refining our sample to adhere to the DESI spectroscopic survey footprint, spanning 14,000 deg$^2$ above a declination of $-18$ degree, yielded a final count of 373,533 galaxies eligible for FP analysis.

\subsection{Spectroscopic Selection}
Our spectroscopic selection process depends on the data obtained from the DESI Survey Validation (SV; \citealt{DESI_Collaboration2023b}). This validation survey unfolded in three phases: firstly, SV1, focused on target selection validation and the refinement of target selection algorithms \citep{Myers2023}; secondly, SV2, referred to as the Operation Development phase, serving as a practice run for the third stage; finally, SV3, also known as the 1\% Survey, aimed to further validate both the survey operations procedures and the final target selection. The culmination of all three stages is encapsulated within the DESI EDR (\citealt{DESI_Collaboration2023}), which has now been released and is publicly accessible.\footnote{https://data.desi.lbl.gov/doc/}

\begin{figure*}
	\includegraphics[width=\textwidth]{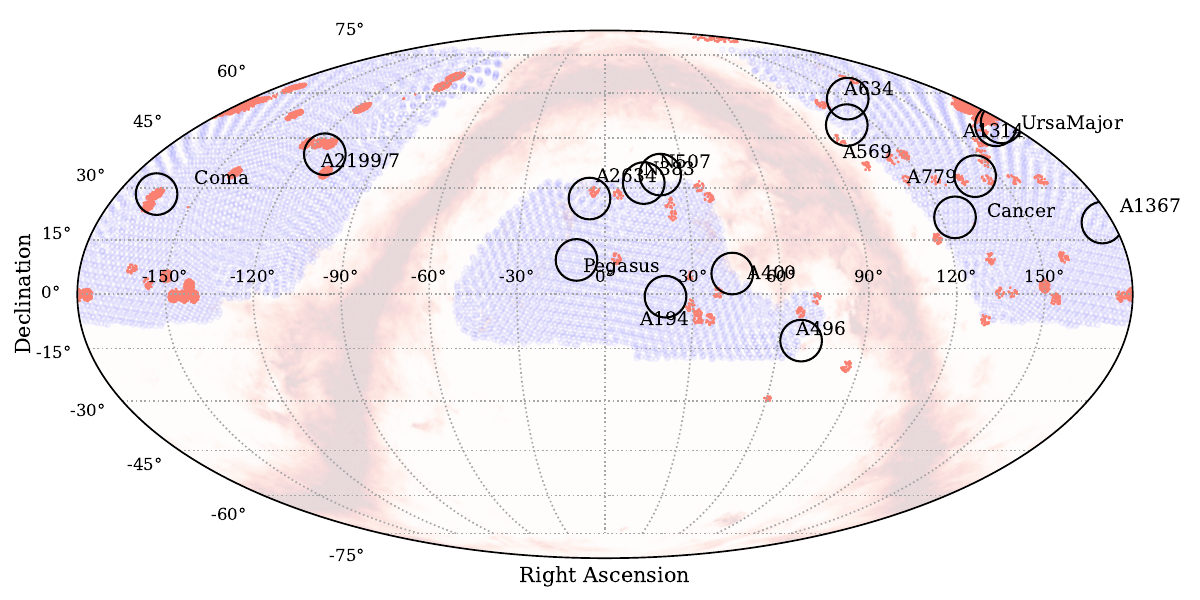}
    \caption{The distribution of FP galaxies within the DESI SV dataset, presented here in a Mollweide projection in equatorial coordinates. Regions obscured by Galactic extinction in the Milky Way are shown as shaded Reds. The orange dots indicate the FP data within the SV dataset. The blue circles encompass all the DESI tiles, which collectively define the DESI survey footprint. Notably, the open black circles pinpoint a few clusters within the DESI footprint, which have previously measured distances \protect\citep{Bell2023}, making them invaluable for the zero-point calibration in the future. For the Python code and data used to reproduce this plot, see \href{https://github.com/KSaid-1/DESI_fuji_FP/tree/main/Plots/Fig.1}{this link.}}
    \label{desi_fuji_aitoff}
\end{figure*}

From our photometrically selected FP sample, we identified the galaxies that were spectroscopically observed in the SV data. We then applied two key constraints to this subset: firstly, the warning bitmask (\texttt{ZWARN}) must be zero, indicating the absence of known issues with the data or the fit; secondly, the spectral classification (\texttt{SPECTYPE}) should be "GALAXY" \citep{DESI_Collaboration2023}. The outcome of this cross-matching process yielded a total count of 6698 distinct FP galaxies. Figure \ref{desi_fuji_aitoff} shows the distribution of these galaxies in a Mollweide projection in equatorial coordinates. The open circles in the figure represent several clusters within the DESI footprint, which are key to the FP analysis, especially for setting the zero-point calibration.

\subsection{Visual inspection}
In order to identify galaxies that were unsuitable for inclusion in our FP analysis, a visual inspection of all galaxies was conducted using $1 \times 1$ arcmin colour cutouts sourced from the Pan-STARRS1 \citep{Chambers2016} and the DESI Legacy Imaging Surveys \citep{Dey2019} images. This visual inspection was carried out by John R. Lucey. Using deeper Legacy Survey images, particularly model residual images, significantly improved our discrimination capability compared to previous surveys. Our methodology followed established procedures from prior FP studies \citep{Campbell2014,Said2020,Howlett2022}.

During the visual inspection, the following categories of objects were identified:
\begin{enumerate}
  \item Galaxies that are not bulge-dominated, including those with prominent spiral arms.
    \item Galaxies for which the measurements of the FP photometric parameters, specifically total magnitude and effective radius, were likely to be unreliable due to the presence of overlapping sources, whether stars or other galaxies.
    \item Galaxies with pronounced central asymmetries, including those with strong dust features, which are likely to bias the velocity dispersion measurements. 
\end{enumerate}
The results of our visual inspection process, are summarized in table \ref{tab:selection criteria}. This visual classification was conducted to assess whether such subjective morphological filtering improves the FP constraints. However, we have tested the impact of these classifications on various galaxy parameters to understand potential biases introduced by visual selection (see section \ref{S_hubble_constant}).

\section{Fundamental Plane Parameters}
\label{S_FP_parameters}
With both the essential photometric and spectroscopic data in hand, we possess the necessary components for deriving the FP parameters. The formulation of the FP relation employed in this study is characterized by its structure:

\begin{eqnarray}
    \log R_e = a \log \sigma_0 + b \log I_e + c.
    \label{FP_equation}
\end{eqnarray}

In this equation, $R_e$ stands for the effective radius, measured in (kpc h$^{-1}$), which serves as the parameter that allows us to measure distance. On the other side of the equation, $\sigma_0$ signifies the central velocity dispersion, expressed in (km s$^{-1}$), while $I_e$ represents the mean surface brightness within the angular effective radius, presented in (L$_{\odot}$~pc$^{-2}$). Notably, $\sigma_0$ and $I_e$ are both distance-independent parameters\footnote{By distance-independent, we mean that these parameters can be measured without needing to know the distance to the object.}. The coefficients of the FP relation are represented by $a$, $b$, and $c$.

The computation of the $r-$band angular effective radius, $\theta_e$ was derived from the $r-$band half-light radius, $r$, as well as the ellipticity components: $\epsilon_1$ and $\epsilon_2$. This relationship is expressed through the following equations:

\begin{eqnarray}
    \theta_e &=& r \sqrt{b/a}\\
    b/a &=& \frac{1.0 - |\epsilon|}{1.0+|\epsilon|}\\
    |\epsilon| &=& \sqrt{\epsilon_1^2 + \epsilon_2^2}.
\end{eqnarray}

In the above equations, $r$, $\epsilon_1$, and $\epsilon_2$ are all directly extracted from the DESI Legacy Imaging Surveys DR9\footnote{https://www.legacysurvey.org/dr9/catalogs/} \citep{Dey2019}.

We converted the angular effective radius in arcseconds to the physical effective radius in units of kpc h$^{-1}$ using the angular diameter distance \citep{Weinberg1972}. This conversion was performed with respect to the observed redshift in the CMB frame, following the standard $\Lambda$CDM cosmological model:

\begin{multline}
    \log R_{e} = \log(\theta_{e}) + \log(d(z_{\mathrm{cmb}})) - \log(1+z_\mathrm{helio})\\ + \log \left(\frac{1000\pi}{180 \times 3600}\right).
    \label{R_e}
\end{multline}

While the comoving distance calculation employed the redshift in the CMB frame, the heliocentric redshift was used for the conversion from comoving to angular diameter distance as recommended by \cite{Davis2019}.

The calculation of the second parameter in the FP relation, the effective surface brightness $I_e$ in L$_{\odot}$~pc$^{-2}$, was based on the model flux in the $r-$band, $f_r$, and Galactic transmission in the same band $\text{MW}_r$. This, along with the above calculated angular effective radius $\theta_e$, was employed in the following manner:

\begin{multline}
\log I_e = 0.4(M_{\odot}^{r} - m_{r} - 0.85z_{\text{cmb}} + k_{r}) - \log(2\pi\theta_{e}^{2}) \\
+ 4\log(1+z_{\text{helio}}) + 2\log(206265/10),
\label{eq:FPi}
\end{multline}
where,
\begin{eqnarray}
    m_r = 22.5 - 2.5\log \frac{f}{\text{MW}_r}. 
\end{eqnarray}

Here, $M_{\odot}^r = 4.65$ signifies the $r-$band absolute magnitude of the Sun \citep{Willmer2018}, while $m_r$ is the extinction-corrected magnitude. The term $0.85z$ accounts for evolution correction \citep{Bernardi2003b}, $4\log(1+z_\text{helio})$ represents surface brightness dimming correction, $k_r$ approximates the $K-$correction in the $r-$band given by \cite{Chilingarian2010}, and $f$ and $\text{MW}_r$ are directly extracted from the Legacy Imaging Surveys DR9.

Completing the set of parameters for the FP, the third parameter is the distance-independent central velocity dispersion, $\sigma_0$. In our analysis, we employed the Penalized Pixel-Fitting (pPXF) software to measure velocity dispersion and its associated uncertainty from DESI spectra. As the name suggests, pPXF utilizes the maximum penalized likelihood method for deriving stellar kinematics from absorption-line spectra of galaxies. The development of this method was initiated by \cite{Cappellari2004} and has been refined in subsequent works by \cite{Cappellari2017,Cappellari2022}. Our stellar templates were drawn from the Indo-U.S. Coudé Feed Spectral Library \citep{Valdes2004}, which encompasses 1273 stars, though typically only about 10-20 are selected by pPXF for precise fits. This spectral library spans the range from 3460 to 9464\AA\/ at a resolution of 1.35\AA\/, corresponding to $\sigma = 30$ km s$^{-1}$. 

The DESI spectrograph's resolving power $(\lambda/\Delta \lambda)$ varies as a function of wavelength: [2000, 3500] in the blue, [3300, 5000] in the red, and [3500, 5200] in the z band (see Fig. 33 in \citealt{DESI_Collaboration2022}). This translates to resolutions for measuring velocity dispersions of 46, 31, and 29 km s$^{-1}$ in the blue, red, and z channels, respectively. In this paper, we use only the blue channel data. The stellar template library we are using here has a higher resolution than the data, which is necessary for accurate fitting. Additionally, the instrumental resolution in pPXF was handled by using the full-resolution DESI data matrix and then outputting a 1D array with the RMS per pixel. 

DESI spectra are categorized into two main groups: full-depth and per-tile spectra \citep{DESI_Collaboration2023}. To derive the final central velocity dispersion, we primarily utilized the full-depth spectra. These spectra merge exposures for targets positioned on a given sky pixel and also aggregate data across tiles when the same target is observed on multiple tiles. On the other hand, per-tile spectra do not combine data from various tiles, even if the same target was observed on multiple tiles. Per-tile spectra find utility in subsequent subsections, where they are used for internal consistency checks.

Subsequently, the measured velocity dispersion was transformed into central velocity dispersion using the formula introduced by \cite{Jorgensen1995}:

\begin{eqnarray}
    \frac{\sigma_0}{\sigma} = \left( \frac{\theta_e/8}{\theta_{\rm ap}}\right)^{-0.04}.
\end{eqnarray}

Here, $\theta_e/8$ correspond to the standard aperture size (one-eighth of the optical effective radius), and $\theta_{\rm ap}=0.75$ represents the DESI fiber radius in arcseconds \citep{DESI_Collaboration2022}. 

Following the computation of the three FP parameters, $R_e$, $I_e$, and $\sigma_0$, we also calculated the corresponding uncertainties, $\epsilon_r$, $\epsilon_i$, and $\epsilon_s$, respectively. These uncertainties are pivotal in accounting for the overall scatter observed in the FP relation. Among these sources of scatter, the intrinsic scatter of the relation itself holds the primary contribution. Nonetheless, the velocity dispersion error also plays a substantial role, ranking as the second most influential factor. If the velocity dispersion error exceeds 20\%, it begins to dominate the scatter, surpassing the intrinsic component.

\begin{figure}
	\includegraphics[width=\columnwidth]{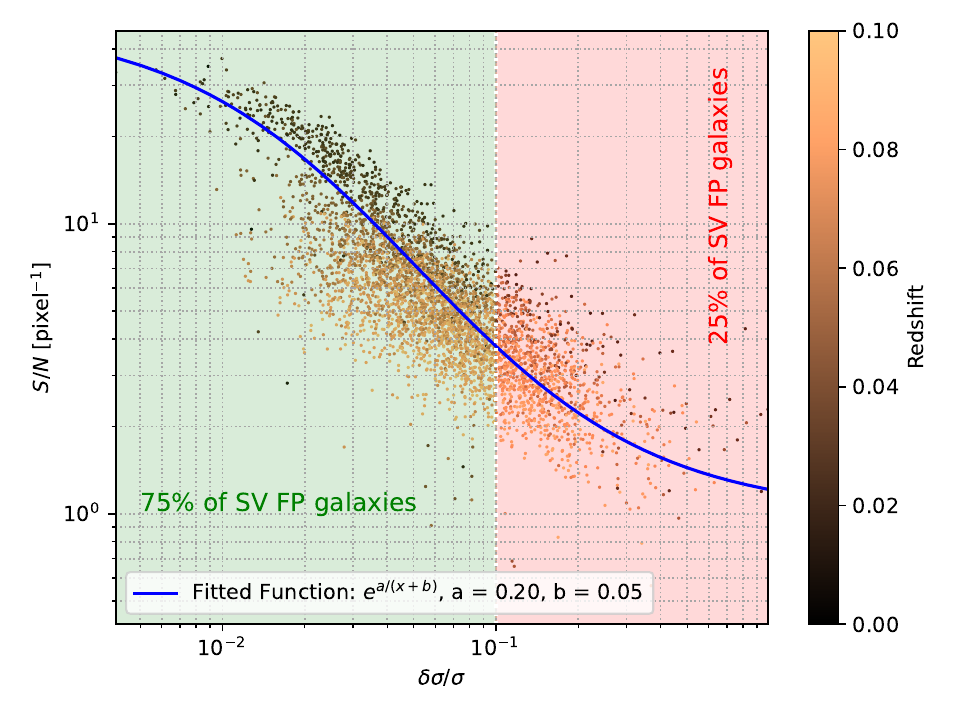}
    \caption{Observed FP galaxies SNR as a function of the relative error in the velocity dispersion measurements from pPXF. Galaxies are color-coded by their redshift. The green-shaded region indicates galaxies with relative errors less than 10\%, while the red-shaded region shows galaxies with errors exceeding 10\%. 75\% of the observed galaxies have relative errors less than 10\%. The fitted curve suggests that even with SNR as low as 4, one can still get a velocity dispersion measurement with less than 10\% relative error. For the Python code and data used to reproduce this plot, see \href{https://github.com/KSaid-1/DESI_fuji_FP/tree/main/Plots/Fig.2}{this link.}}
    \label{SNR_vs_delta_sigma}
\end{figure}

Figure \ref{SNR_vs_delta_sigma} illustrates the Signal-to-Noise Ratio (SNR) of the observed FP galaxies in the blue band against the relative error in velocity dispersion measurements derived from pPXF. 
In this figure, galaxies are colour-coded according to their redshift. The SNR calculated is per-pixel, representing the median signal divided by the standard deviation of the residuals. Notably, the figure showcases that approximately 75\% of the observed galaxies exhibit relative errors of less than 10\%, rendering them suitable candidates for FP.

\subsection{Internal consistency}
To perform an internal consistency assessment of our DESI FP sample, we will employ the entire set of 6698 galaxies, encompassing the full FP sample prior to implementing the redshift and velocity dispersion constraints. Within this subsection, we will systematically conduct an in-depth internal consistency examination of the FP parameters, both photometric and spectroscopic.

 In our FP analysis, we utilized data from the DESI Legacy Imaging Surveys (LS; \citealt{Dey2019}) to derive all our photometric parameters. 

It is important to note that due to the utilization of various telescopes, cameras, and filters combinations in the BASS and DECam LS, systematic variations in the zero-point calibration between these two photometric systems are observed (as detailed in \citealt{Dey2019}, section 7.2).

To investigate any systematic difference, we conducted a comparison of the 5 arcsec aperture magnitudes for galaxies in our FP sample that were observed in both the BASS and DECaLS surveys.

Among the 1660 galaxies shared between the two surveys, we observed a median BASS $-$ DECaLS difference of $+0.0234$ mag, with root mean square (RMS) deviation of $0.02$ mag. In light of this, we incorporate this correction into our analysis by adjusting the northern $r-$band magnitudes, by $0.0234$ mag. This offset will be investigated further with year 1 data, which will include approximately 100,000 galaxies, providing a more robust calibration for our FP data

For the velocity dispersion measurements, we used data from the DESI spectroscopic survey early data release \citep{DESI_Collaboration2023}. In contrast to the employment of full-depth spectra for the final determination of central velocity dispersion, we utilize per-tile spectra for our spectroscopic data in this context. Per-tile spectra prove to be particularly valuable for the internal consistency assessment of our velocity dispersion measurements. These spectra combine observations across multiple exposures within the a single tile, but not across different tiles. This allows us to treat per-tile spectra as repeated observations of the same targets. By comparing velocity dispersion from the same target observed on different tiles, we can identify potential tile-to-tile offsets due to varying observing conditions.

For each of our FP galaxies, our initial step involved determining whether they had been observed on different tiles more than once. Subsequently, we categorized these observations pairs as primary or secondary based on the signal-to-noise, with observations having the higher SNR being considered the primary one. After this categorization, we proceeded to measure the velocity dispersion for all pairs, employing the same methodology applied to the full-depth spectra, as explained earlier. 

In order to assess the consistency of the velocity dispersion measurements and explore the possibility of systematic offsets between observations or tiles within the DESI data, we employed the relative error between pairs of observations. This relative error encompasses both the measurements of the velocity dispersion itself and its associated error. The assessment was conducted using the following formula:

\begin{eqnarray}
    \epsilon = \frac{\sigma_p-\sigma_s}{(\delta \sigma_p^2 + \delta \sigma_s^2)^\frac{1}{2}}.
\end{eqnarray}

Here, $\sigma_p$, $\sigma_s$, $\delta \sigma_p$, and $\delta \sigma_s$ represent the velocity dispersion measurements from primary and secondary tiles, accompanied by their respective error estimates. In the context of consistent and unbiased velocity dispersion measurements, with accurately estimated errors, this evaluation should yield a Gaussian distribution with a mean of zero and a standard deviation of unity.

\begin{figure*}
	\includegraphics[width=\textwidth]{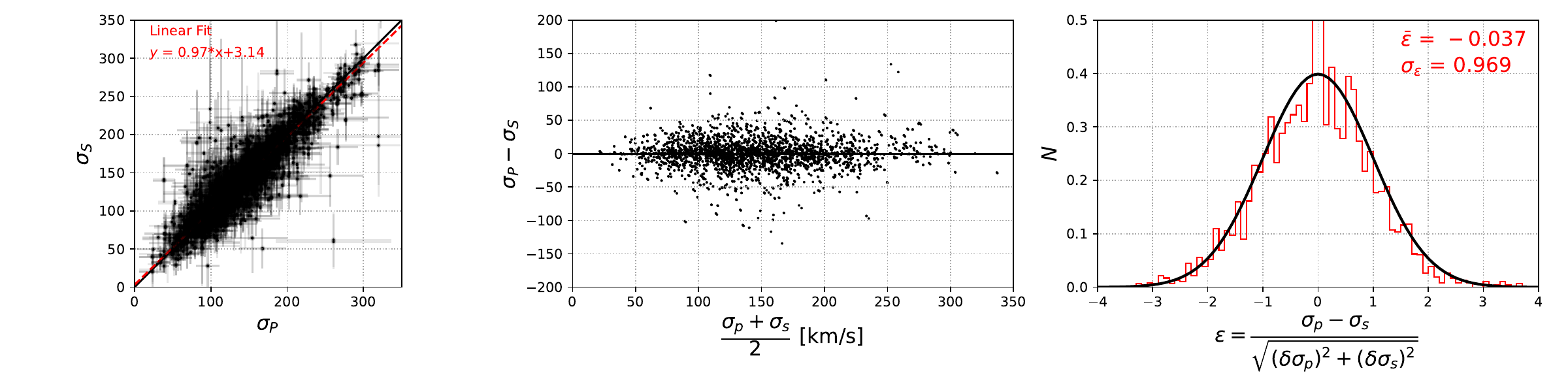}
    \caption{Pairwise comparison of velocity dispersion measured using pPXF in this work between repeat observations for a subset of DESI galaxies selected for FP relation. Left Panel: one to one comparison between primary and secondary observations where primary is defined as the observations with the highest SNR. The red line shows the linear fit result. Middle Panel: the difference between the two measurements as a function of the mean of the two measurements. Right Panel: distribution of pairwise relative errors in velocity dispersion measurements. The distribution is in a good agreement with a Gaussian with mean of zero and standard deviation of unity shown as a solid curve. For the Python code and data used to reproduce this plot, see \href{https://github.com/KSaid-1/DESI_fuji_FP/tree/main/Plots/Fig.3}{this link.}}
    \label{internal_consistency}
\end{figure*}
In Figure \ref{internal_consistency}, we present an assessment of the velocity dispersion measurements conducted using pPXF and the Indo-US. stellar library. This evaluation is carried out on a subset of 4644 pairs from a total of 1420 unique galaxies in our DESI FP sample, utilizing per-tile DESI spectra. The left panel of the figure illustrates the one-to-one comparison between primary and secondary tiles. Reassuringly, the observations exhibit agreement within the uncertainties for the majority of tiles.

For enhanced clarity regarding any differences, we plot the difference between the measurements of each pair as a function of their mean in the middle panel. To provide a quantitative analysis of these differences, we construct a histogram in the right panel to visualize the distribution of pairwise relative errors in velocity dispersion measurements. The solid curve in this panel represents a Gaussian distribution with a mean of zero and a standard deviation of one. Notably, this figure affirms that the measurements are consistent and unbiased, reinforcing the reliability of our velocity dispersion measurements.

\subsection{External consistency}
With a successful internal assessment of our velocity dispersion measurements across different tiles and observing conditions, we now embark on a more extensive evaluation by comparing our measurements to those obtained by other surveys and telescopes. Our objective is to ensure that our measurements align closely with those from well-vetted sources. To achieve this, we perform a comparison between our velocity dispersion measurements derived from the full-depth spectra and those from the Sloan Digital Sky Survey (SDSS).

In this endeavor, we cross-match the entire FP sample of 6698 galaxies, prior to any spectroscopic cuts, with the SDSS Data Release 14 \citep{Abolfathi2018}. Our cross-match yields 4221 galaxies that are present in both the DESI FP sample and the SDSS survey. To assess the agreement, we employ the same methodology of relative error comparison that was used for the internal consistency check. 

However, it is important to note that SDSS provides two distinct velocity dispersion values: firstly, from the base table containing all spectroscopic information, known as \texttt{veldisp} (referred to as the pipeline value); secondly, from the emissionLinesPort catalogue, from the Portsmouth group, which utilizes the pPXF method and the MILES stellar library \citep{Sanchez-Blazquez2006} for stellar kinematics measurements, including velocity dispersion, denoted as \texttt{sigmastar} in the SDSS dataset. While there are other SDSS catalogs available for velocity dispersion, comparing two measurements from the same survey is sufficient for our purposes of external consistency checking.

\begin{figure*}
	\includegraphics[width=\textwidth]{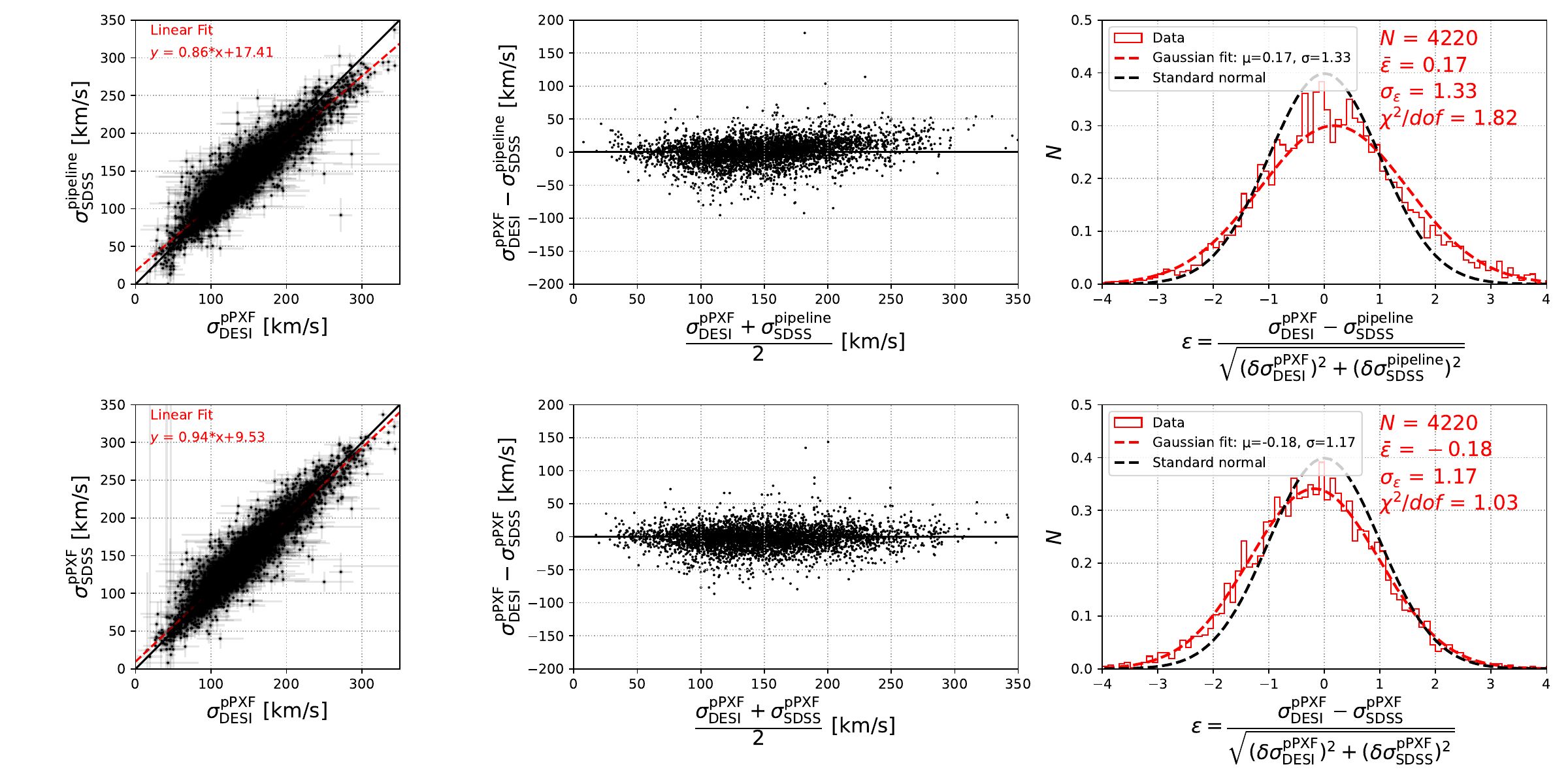}
    \caption{Velocity dispersion measured with pPXF in this work for FP galaxies in DESI SV data in comparison with SDSS measurements. The top panel shows a comparison with the SDSS pipeline velocity dispersion $\sigma_{\rm SDSS}^{\rm pipeline}$. The bottom panel present a comparison with the Portsmouth group (emissionLinesPort) velocity dispersion using pPXF $\sigma_{\rm SDSS}^{\rm pPXF}$. We include quantitative Gaussian fit statistics with $\chi^2$/dof values of 1.82 for DESI pPXF vs. SDSS pipeline (top right) and 1.03 for DESI pPXF vs. SDSS pPXF (bottom right), indicating better statistical consistency between the two pPXF implementations than between different measurement techniques. The agreement between DESI velocity dispersion and SDSS in general is better than the agreement between SDSS pipeline velocity dispersion measurements and the SDSS pPXF ones. For the Python code and data used to reproduce this plot, see \href{https://github.com/KSaid-1/DESI_fuji_FP/tree/main/Plots/Fig.4}{this link.}}
    \label{external_consistency_v3}
\end{figure*}

Figure \ref{external_consistency_v3} presents the outcomes of our comparison. In the top panel, we compare our results to the SDSS pipeline measurements. Notably, there exists an overall agreement between the two measurements. However, a small yet significant offset of 0.17 is observed, along with a standard deviation of approximately 1.3. This offset is around 7 times the standard error in the mean, and the deviation from unity of the standard deviation is 32 times the uncertainty in the standard deviation, highlighting its high significance. 

Given these findings, we extend our comparison to the SDSS pPXF measurements, as displayed in the bottom panel. Interestingly, a comparable offset is observed, but with the opposite sign of -0.18. Importantly, the deviation from unity in the standard deviation is reduced to 1.17.

To quantitatively assess the statistical consistency of our velocity dispersion measurements, we performed chi-squared goodness-of-fit tests on the normalized pull distributions. For the comparison between DESI pPXF and SDSS pipeline measurements, we find $\chi^2$/dof = 1.82, indicating moderate deviation from the expected normal distribution. In contrast, the comparison between DESI pPXF and SDSS pPXF implementations yields $\chi^2$/dof = 1.03, extremely close to the ideal value of 1.0, suggesting excellent statistical consistency. This reveals that systematic differences are minimised when similar measurement techniques are employed, while comparisons between fundamentally different methods (pipeline vs. pPXF) show statistically significant discrepancies even when measuring the same physical property.

As part of a comprehensive approach to assess potential systematic effects, we calibrated our DESI velocity dispersion measurements against the SDSS pPXF measurements using the fitted red line (both slope and intercept) in Figure \ref{external_consistency_v3}. This calibration was done only to evaluate the impact of any systematic offset or tilt if incorrect velocity dispersions were used. Our fiducial cosmological analysis relies on the DESI measurements. The calibration with SDSS was included solely to identify and account for this potential source of systematic bias, which is incorporated into the total error budget, as discussed in Section \ref{S_hubble_constant}.

\section{Fundamental Plane fits}
\label{S_FP_fit}
To derive distances and peculiar velocities, we employed the Maximum Likelihood method to fit the FP using a 3D Gaussian model. This approach, initially formulated by \cite{Saglia2001} and \cite{Colless2001}, has since been refined and adapted by subsequent studies such as \cite{Magoulas2012,Springob2014,Said2020,Howlett2022}. Here, we provide a brief summary of the method, but for a comprehensive review, see the aforementioned studies.

In this study, we employ the conventional two-step approach. First, we perform a FP fitting without considering peculiar velocities. Then, the deviation from the optimal fit of the FP is attributed to the peculiar velocities of individual galaxies. 

We initiate by establishing the core parameters of the FP as $r = \log R_e$, $s = \log \sigma_0$, and $i=\log I_e$. Following the approach proposed by \cite{Colless2001}, we describe the three-dimensional probability distribution in the $(r, s, i)$ space as follows:

\begin{eqnarray}
  P(x_n) = \frac{\exp\left[-0.5\mathbf{x}_n^T (\mathbf{V} + \mathbf{E}_n)^{-1} \mathbf{x}_n\right]}
{(2\pi)^{3/2}|\mathbf{V}+\mathbf{E}_n|^{1/2}f_n}.  
\end{eqnarray}

Here, $\mathbf{x}_n = (r-\bar{r},s-\bar{s},i-\bar{\imath})$ signifies the position of galaxy $n$ within the FP domain. The matrix $\mathbf{V}$ encapsulates the intrinsic scatter of the FP relation, while $\mathbf{E}_n$ accounts for the measurement uncertainties associated with the FP parameters (see equation 14 by \cite{Said2020} for the full description of the error matrix). The normalization factor $f_n$ ensures the distribution integrates to unity, accounting for selection criteria. 

Subsequently, the likelihood can be expressed as:

\begin{eqnarray}
    \Lagr = \prod_{n=1}^{N_g} P(x_n)^{1/S_n}.
\end{eqnarray}

In this context, $S_n$ signifies the $1/V_{\text{max}}$ weighting factor, which accommodates for galaxies that might be absent due to the selection function \citep{Said2020}. The objective is to determine the optimal parameters ($a$, $b$, $\bar{r}$, $\bar{s}$, $\bar{\imath}$, $\mathbf{V}$) of the FP that best describe the data, with $\mathbf{V}$ encompassing the scatter intrinsic to each orthogonal direction, namely $\sigma_{1}$, $\sigma_{2}$, and $\sigma_{3}$.

In our fitting procedure, the parameters $f_n$ and $S_n$ accommodate for three specific selection criteria: (1) a range of lower and upper limits on the $r-$band magnitude, set at $10 \leq m_r \leq 18.$; (2) lower and upper boundaries on redshift, restricted to $0.003 \leq z \leq 0.1$; (3) velocity dispersion, constrained by the instrumental resolution of the DESI spectrographs, adhering to $50 \leq \sigma \leq 420$ km s$^{-1}$. Table \ref{tab:selection criteria} shows the number of remaining galaxies after each successive selection criterion.

\begin{table}
    \centering
    \begin{tabular}{ccc}
    \hline\hline
        Selection Criteria & Number & After Visual inspection\\
     \hline
       No cuts  & 6698 & 4682\\
       $10 \leq m_r \leq 18$  & 6698 & 4682\\
       $0.003 \leq z \leq 0.1$  & 4290 & 3198\\
       $50 \leq \sigma \leq 420$ km s$^{-1}$  & 4191 & 3110\\
        \hline
    \end{tabular}
    \caption{Summary of the DESI FP peculiar velocity Selection criteria}
    \label{tab:selection criteria}
\end{table}

\begin{figure}
	\includegraphics[width=\columnwidth]{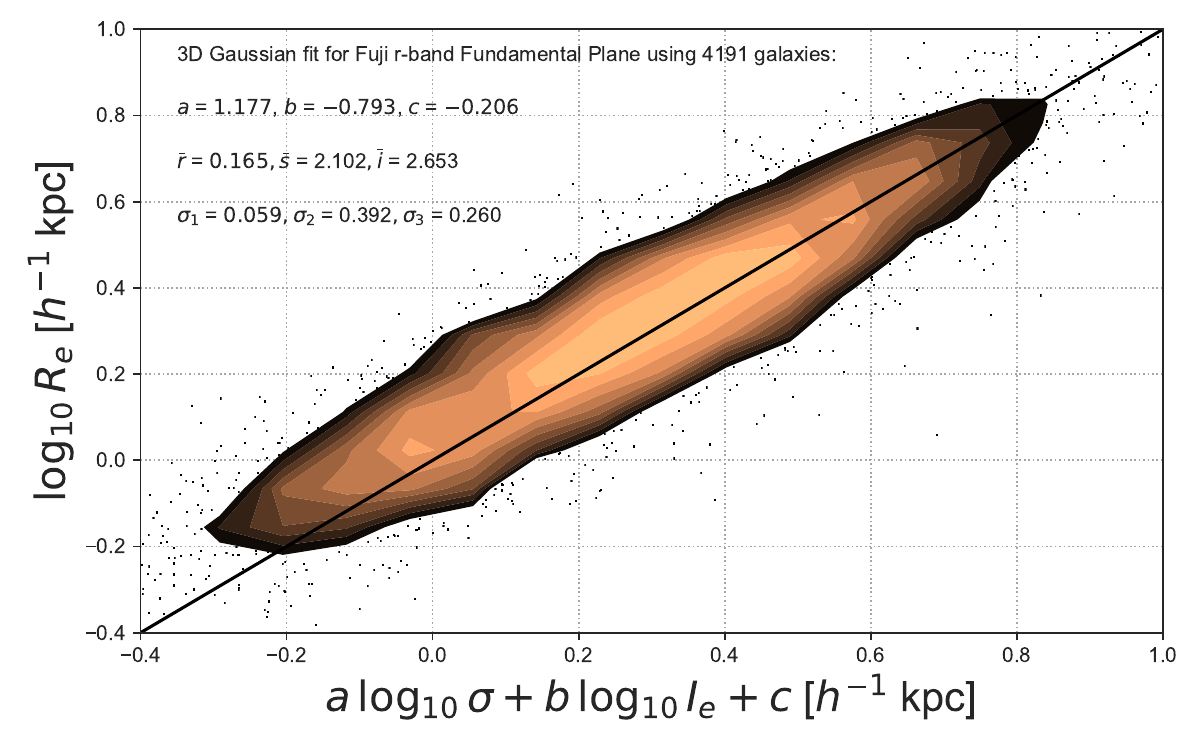}
    \caption{The projected FP of DESI SV data. The data shows the measured effective radii against the predicted radii from the 3D Gaussian fit of the FP for DESI SV data. The solid black line shows the one-to-to line. For the Python code and data used to reproduce this plot, see \href{https://github.com/KSaid-1/DESI_fuji_FP/tree/main/Plots/Fig.5}{this link.}}
    \label{FP_fuji}
\end{figure}

Figure \ref{FP_fuji} illustrates the forward projection of the FP relation, accompanied by the best fit of the FP parameters for the DESI $r-$band sample as derived through this procedure. The FP parameters fitted under different conditions are listed in Table \ref{tab:FP}.

\begin{table*}
\centering
\caption{FP parameters for the DESI SV sample under different conditions: Fiducial (original) and calibrated DESI velocity dispersion measurements to the SDSS measurements from the Portsmouth group using pPXF. SDSS and 6dFGSv FP parameters are also included for comparison.}
\begin{tabular}{lccccc} \hline
Parameter & Fiducial & Calibrating DESI $\sigma$ to SDSS & SDSS \citep{Howlett2022} & 6dFGSv \citep{Magoulas2012}\\ \hline
$N_{\mathrm{gal}}$ & 3110 & 3110 & 34059 & 8803 \\
$a$ & 1.177 & 1.312 & $1.274 $ & 1.523   \\
$b$ & $-$0.793 & $-$0.794 & $-0.841 $ & $-$0.885  \\
$\bar{r}$ & 0.165 & 0.168 & $0.161 $ & 0.184   \\
$\bar{s}$ & 2.102 & 2.119 & $2.174 $ & 2.188  \\
$\bar{i}$ & 2.653 & 2.655 & $2.688 $ & 3.188   \\
$\sigma_{1}$ & 0.059 & 0.056 & $0.054 $ & 0.053  \\
$\sigma_{2}$ & 0.392 & 0.392 & $0.335 $ & 0.318  \\
$\sigma_{3}$ & 0.256 & 0.242 & $0.219 $ & 0.170 
\vspace{1pt} \\ \hline
\end{tabular}
\label{tab:FP}
\end{table*}

With the complete set of FP parameters at our hands, we are positioned to compare them with findings from previous works.  Notably, as the FP parameters are contingent upon wavelength, a straightforward method for comparison and assessment against earlier studies involves the computation of the root mean square (RMS) scatter of the FP in the $r$ direction. This measure should offer valuable insights into the actual distance error. However, it is worth emphasizing that the true distance error encompasses supplementary elements, including the correction for the selection function and the underlying distribution of galaxies within the FP, and the full covariance between parameters such as $I_e$ and $\theta_e$. While we present a simplified estimate here, our full analysis incorporates these complex relationships through the use of complete covariance matrices.

The comprehensive RMS scatter in the $r$ direction can be quantified through the expression:

\begin{eqnarray}
    \sigma_r = \left[ (a\epsilon_s)^2 + \epsilon_{\rm phot}^2 + \sigma_{r,{\rm int}}^2 \right]^{1/2}.
    \label{sigma_r}
\end{eqnarray}

In this equation $\epsilon_s$ denotes the mean error in $\log\sigma$, while $\epsilon_{\rm phot}$ is the total photometric error arising from both $\epsilon_r$ and $\epsilon_i$, determined as $\epsilon_{\rm phot} = \left[ \epsilon_r^2 + b\epsilon_i^2 \right]^{1/2}$. Additionally, $\sigma_{r,{\rm int}}$ signifies the intrinsic scatter within the FP itself. Upon substituting these values into equation \ref{sigma_r}, the total RMS scatter in the $r$ direction is established as 24.4\%. This outcome marks a notable improvement compared to the 6dFGSv reported value of 31\% \citep{Magoulas2012}. It is worth noting that while this value is similar to the total scatter evident in the SDSS FP \citep{Said2020}, the DESI number of FP galaxies is projected to be at least five-fold greater than the SDSS FP sample, which itself stands as the most expansive peculiar velocity survey undertaken thus far.

That was essentially the initial step in the traditional two-step FP approach. Up to this point, we have worked under the assumption of negligible peculiar velocities. However, the next phase involves calculating these peculiar velocities by gauging the deviations of the data from the best-fit FP parameters. By comparing the physical effective radius, $r$, derived from equation \ref{R_e}, with the true effective radius, $r_t$, inferred from the best-fit FP parameters, we can derive the log-distance ratio as:
\begin{eqnarray}
    r  - r_{t} = \eta.
\end{eqnarray}
The log-distance ratio is the main derived value in our dataset.

\section{Zero-point calibration and absolute distances}
\label{S_zero_point}
In the FP equation \ref{FP_equation}, the coefficient $c$ sets the zero-point of the relation. The determination of this zero-point has a direct impact on the calculation of the actual effective radius of galaxies, subsequently influencing the derived distances and peculiar velocities. When fitting the FP, one approach is to assume that the net radial peculiar velocity of all galaxies is zero, which corresponds to assuming no monopole term in the velocity field. However, this assumption is problematic for hemispheric surveys such as 6dFGSv, SDSS, and DESI, as it requires full sky coverage. Therefore, a proper zero-point calibration becomes essential.

Various surveys have adopted different approaches to establish the zero-point. For instance, \cite{Springob2014} utilized a sub-sample of the 6dFGSv near the celestial equator, defining a great circle sample to re-fit the FP and adjust the zero-point for the entire sample. This effectively treats the sample as a full-sphere, being degenerate only in the monopole term. Unfortunately, this approach is not feasible for the DESI early data release due to the sky coverage limitations. However, it holds promise for the full DESI data release.

Another example can be found in the SDSS peculiar velocity survey by \cite{Howlett2022}. In this study, they performed a cross-match between the SDSS sample and the Cosmicflows-III catalogue (CF3; \citealt{Tully2016}), which contains distance measurements from alternative methods. 

We adopted an approach similar to SDSS's method by cross-matching our DESI FP sample with the SDSS peculiar velocity catalogue. Specifically, we utilized their calibrated log distance ratios, which were originally calibrated using the CF3 data sets. This dataset itself was calibrated through a distance ladder approach encompassing various standard candles, such as Cepheid variables, Tip of the Red Giant Branch, and Type Ia supernovae. We identified 896 galaxies common to both samples, providing a substantial number of galaxies for zero-point calibration.

\begin{figure}
	\includegraphics[width=\columnwidth]{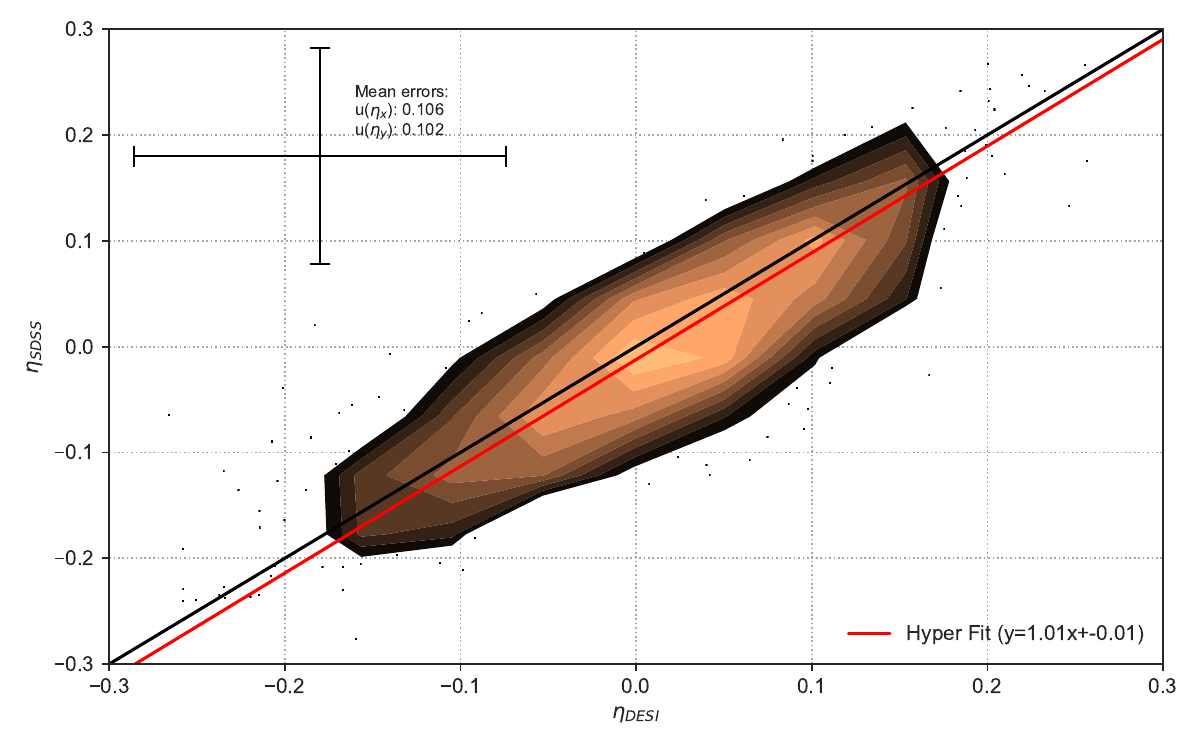}
    \caption{Comparison between the measured log-distance ratio for DESI SV galaxies and their counterpart in the SDSS PV. The solid line shows the one-to-one line. The red line shows the Hyper Fit result (y = 1.01x - 0.01). Mean uncertainties for $\eta_\text{DESI}$ (x-axis) and $\eta_\text{SDSS}$ (y-axis) are displayed in the upper left corner, demonstrating that the observed offsets between methods are statistically insignificant. For the Python code and data used to reproduce this plot, see \href{https://github.com/KSaid-1/DESI_fuji_FP/tree/main/Plots/Fig.6}{this link.}}
    \label{fuji_vs_sdss_logdist}
\end{figure}

Figure \ref{fuji_vs_sdss_logdist} shows a comparison of the log distance ratios for these common objects between SDSS and DESI peculiar velocity catalogue prior to zero-point calibration.

A one-to-one line (black) and a Hyper Fit line (red) are overlaid to show the difference between $\eta_\text{DESI}$ and $\eta_\text{SDSS}$. The large error bars displayed on the top corner reflect the log distance ratio uncertainties derived from the FP, showing that the offset between methods is small compared to these measurement uncertainties. 

However, this approach is not the optimal choice if the desired scientific goal is to measure the present-day expansion rate $H_0$, as this value is already set to $75\pm2$ km s$^{-1}$ Mpc$^{-1}$ for the CF3 catalogue. To address this, we employed an alternative method to calibrate the zero-point (which is degenerate with the Hubble constant $H_0$) using measured distances to clusters in our sample, leveraging primary distance indicators and offering absolute distance measurements. This approach offers two key advantages: First, it grants us greater control over the selection of primary distance indicators, enabling us to choose and exclude indicators as needed; second, by averaging over multiple distances within the same cluster, we can mitigate the overall scatter in the calibration process, in contrast to using individual galaxy distances. 

Although there is not a group/cluster catalogue available for the DESI early data release yet, we know that this release has already covered the Coma cluster to a greater depth than any survey before. The Coma cluster, being massive, relatively nearby, and extensively studied, serves as an ideal candidate for setting the FP zero-point. To make use of additional galaxies in the Coma cluster that had not previously been observed before DESI, we adapted a modified friends-of-friends (FoF) algorithm based on the methodology introduced by \cite{Press:1982}. We combined redshift data from the DESI early data release and supplemented it with SDSS redshifts for galaxies not observed by DESI. Our initial dataset encompassed all DESI and SDSS redshifts within a generous region around the Coma cluster, spanning a 10 degree radius centered on the Coma cluster RA=12h59'48.7", Dec=27\textdegree 58'50" and redshift range of 0.02 relative to the mean redshift of the Coma cluster ($z=0.0231$, \citealt{Abell:1989}). We applied a modified FoF algorithm, implementing separate linking lengths for the angular and radial directions, as outlined in \cite{Eke:2004a,Duarte:2014,Duarte:2015}, and optimized these lengths using a cost function \citep{Robotham:2011} along with SDSS-like mock catalogues. Specifically, we employed linking lengths of 600 km s$^{-1}$ in the radial direction and 0.3{\textdegree} in the angular direction. This yielded a catalogue of 1731 galaxies potentially belonging to the Coma cluster. 

Subsequently, we employed the method of \cite{Jaffe:2015} to remove galaxies that could not be kinematically bound to the Coma cluster, and we excluded galaxies located beyond the cluster's turn-around radius, following \cite{Korkidis:2020}. This process yielded a final catalogue of 1696 identified Coma cluster members. To validate our method and the chosen linking lengths, we applied the same process to DESI-only and SDSS-only datasets. The results were consistent with each other, differing by only a few galaxies. A more detailed description of this method and the associated code for cluster member identification will be provided in Saulder et al., (in preparation).

Cross-matching our FP sample with this newly created Coma catalogue, we identified 226 galaxies in common. We then calculated the uncalibrated distance of the Coma cluster using the weighted mean of distances for these 226 galaxies. To perform zero-point calibration and obtain absolute distances for our full sample, we compared this calculated Coma distance to the most recent absolute distance measurement of the massive NGC 4874 galaxy at the core of the Coma cluster, determined using the Surface Brightness Fluctuation method (SBF; \citealt{Jensen2021}), which yielded $99.1\pm5.8$ Mpc.

We applied this correction between the uncalibrated distance of Coma obtained from the FP and the measurement of Coma's distance from the surface brightness fluctuation method to obtain absolute distances for all galaxies in our sample. 

As with the previous methods, this approach also comes with its own set of limitations. Firstly, due to the absence of a dedicated full group catalogue for DESI data at present, we are confined to utilizing the Coma cluster alone. While this provides a valuable calibration point, it could potentially introduce bias compared to using multiple clusters and groups that span a broader region of the sky. However, it's worth noting that this limitation will be significantly alleviated in future DESI data releases and its associated group catalogue, which will encompass numerous clusters and groups with known distances, facilitating a more comprehensive zero-point calibration process. In Figure~\ref{desi_fuji_aitoff}, we display a selection of these clusters and groups, each possessing known absolute distances, that are well-suited for the calibration procedure. 

The second limitation arises from the observation by \cite{Howlett2022} of a correlation between group richness and one of the FP parameters, specifically the mean surface brightness ($I_e$). Consequently, relying solely on the Coma cluster for zero-point calibration might introduce bias into our results. However, this concern is mitigated by the potential approach of employing multiple FP fits as a function of group/cluster richness, as suggested by \cite{Howlett2022}. In general, while the zero-point calibration can be executed through various methodologies, the choice is contingent on the data available for analysis.

In summary, the process of establishing the zero-point for the FP involves several considerations, each with its own merits and limitations. The methods applied depend on the available data and the specific scientific objectives of the study. With forthcoming DESI data, these challenges will be further addressed and the calibration process can be refined, resulting in even more accurate distance measurements and peculiar velocities for a much larger sample of galaxies. 

\section{Measuring the Hubble Constant}
\label{S_hubble_constant}
The recession velocity-distance relation, often referred to as the Hubble-Lema\^itre law, establishes the connection between a galaxy's  recession velocity and its distance. It is expressed as: 
\begin{eqnarray}
    v = H_0 D.
\end{eqnarray}
In this equation, the constant of proportionality, denoted as $H_0$, represents the present rate of expansion of the universe, Hubble's constant. Our method of measuring the Hubble constant here involves the construction of a Hubble diagram, which necessitates a dataset presenting the distance modulus versus redshift.

However, our approach to creating the Hubble diagram and performing the fitting to estimate the Hubble constant differs from the traditional Hubble diagram. Instead of using apparent magnitude and absolute magnitude as the observable and measured quantities, we employ a different set of variables in the FP analysis. Specifically, our observable quantity is the angular effective radius, while the measured quantity is the true physical radius. Consequently, rather than plotting the distance modulus on the y-axis of the Hubble diagram, our y-axis represents the difference between the logarithm of the angular effective radius and the logarithm of the true physical radius. It is mathematically expressed as:

\begin{eqnarray}
    \mu_{A} &=& r_t - r_\theta\\
           &=& r - \eta - r_\theta - \log (\frac{1000\pi}{180 \times 3600})\\
           &=& \log d_A
\end{eqnarray}          
where $d_A$ is the angular diameter distance parameterized as follows:

\begin{eqnarray}
    d_A &=& \frac{d_L}{(1+z)^2}\\
        &=& \frac{c}{(1+z)} \int_0^z \frac{dz'}{H(z')}
\end{eqnarray}
where $d_L$ is the luminosity distance.
We then express the angular distance modulus as a power series of the form\footnote{This equation is the power series expansion of the angular distance modulus in terms of redshift $z$, a standard approximation in cosmology for low to moderate redshifts. This approximation is accurate for $z<<1$, which covers the range of our data \citep{Visser2004,Weinberg2008}.}:
\begin{multline}
    \mu_{A}^{\text{model}} = \log cz \left(1 + \frac{1}{2} [1 - q_0]z - \frac{1}{6} [1 - q_0 - 3q_0^2 + j_0]z^2\right) \\
           - \log H_0 - 2 \log (1+z).
           \label{k_modulus}
\end{multline}

One can then use this formulation to fit for the Hubble constant $H_0$, the deceleration parameter $q_0$, and the jerk parameter $j_0$ by measuring the angular distance modulus as a function of redshift up to terms of order $z^3$.

\begin{figure*}
	\includegraphics[width=\textwidth]{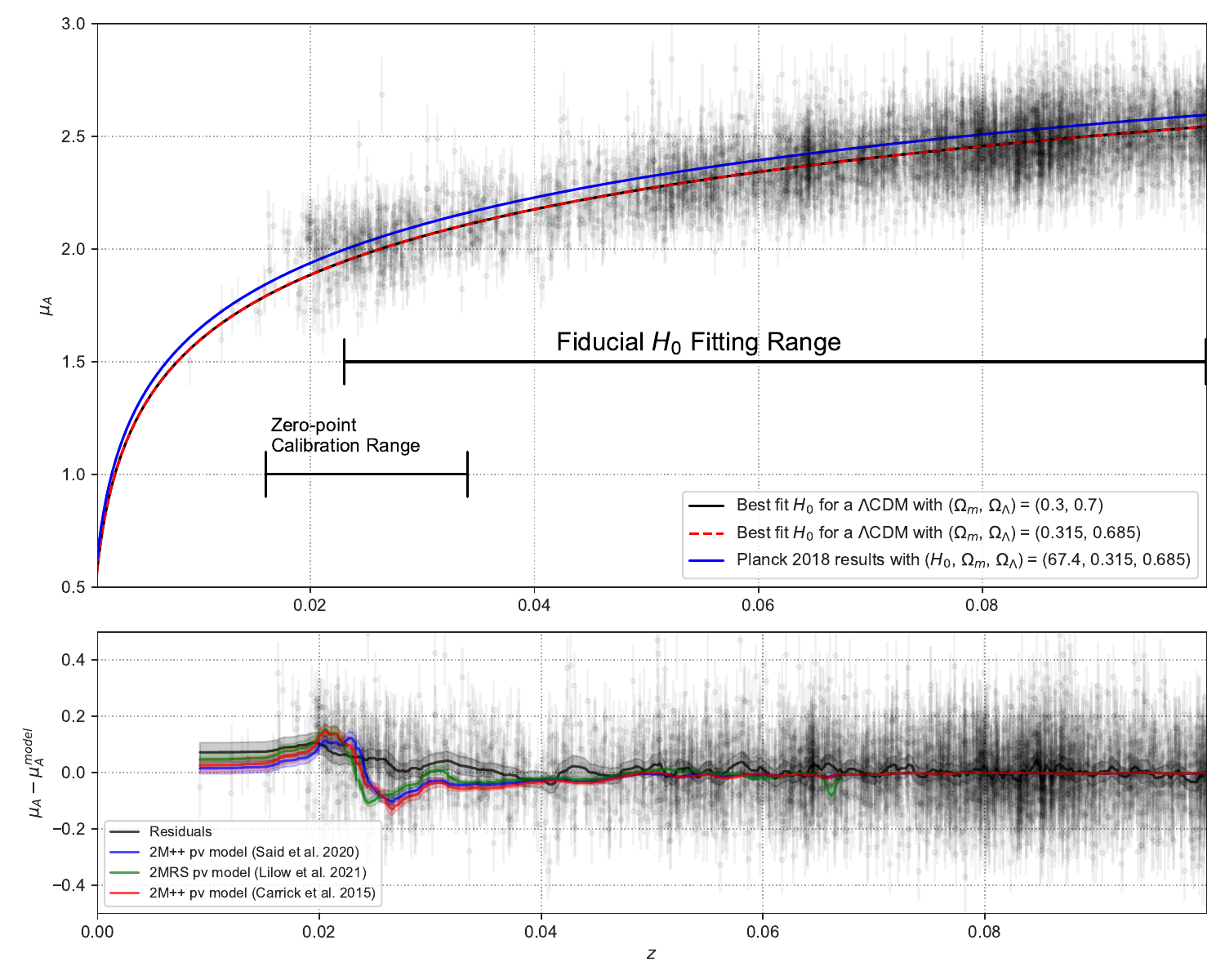}
    \caption{The DESI Hubble diagram features 4191 FP galaxies with redshifts falling within the range of 0.01 to 0.1. In the upper panel, the plot displays the log of the angular diameter distances, denoted as $\mu_{A}$, as a function of redshift, $z$. The solid black curve corresponds to the best-fit $H_0$ for a $\Lambda$CDM cosmology with $\Omega_m = 0.3$ and $\Omega_{\Lambda}=0.7$, as determined by employing equation \ref{k_modulus}. The highlighted regions indicate the redshift ranges used for zero-point calibration and for fitting cosmological parameters. The red dashed line shows our best-fit $H_0$ using Planck values for $\Omega_m$ and $\Omega_{\Lambda}$. The blue curve represent Planck's values for $H_0$, $\Omega_m$ and $\Omega_{\Lambda}$.  In the lower panel, the plot reveals the residuals relative to the best-fit model, calculated as $\mu_A - \mu_{A}^{\text{model}}$. Additionally, in the bottom panel, a black line represents the average trend of the residuals. The observed larger $\mu_A$ at redshift $z<0.023$ is due to LSS and can be explained by a single attractor model showing infall and backside infall toward the attractor's centre. The plotted peculiar velocity field reconstructions 2M++ from \protect\cite{Said2020} and \protect\cite{Carrick2015} and 2MRS from \protect\cite{Lilow2021} also show this trend. For the Python code and data used to reproduce this plot, see \href{https://github.com/KSaid-1/DESI_fuji_FP/tree/main/Plots/Fig.7}{this link.}}
    \label{Hubble_diagram}
\end{figure*}

DESI FP Hubble diagram comprising 4191 FP galaxies within the redshift range of 0.01 to 0.1 is shown in the upper panel of Figure \ref{Hubble_diagram}. 

Due to the limited redshift range covered by our FP data ($z<0.1$), it is not feasible to perform a concurrent fit for all three cosmological parameters in equation \ref{k_modulus}. Instead, we adopt $q_0 = -0.55$ and $j_0=1$, in line with the expectations for a flat $\Lambda$CDM cosmology with $\Omega_m=0.3$ and $\Omega_{\Lambda}=0.7$. We solely perform a fit for $H_0$ under these conditions.

Additionally, following the approach by \cite{Riess2022}, we implement an additional redshift cut of $z>0.023$, to limit the effect of peculiar velocities, which results in a sample of 4063 galaxies. After applying these criteria, the derived value for the Hubble constant is $H_0 = 76.05\pm0.35$ $\mathrm{km \ s^{-1} Mpc^{-1}}$. This uncertainty exclusively encompasses statistical uncertainties. 

Figure \ref{Hubble_diagram} illustrates our best-fit model for $H_0$ using a flat $\Lambda$CDM cosmology with $\Omega_m=0.3$ and $\Omega_{\Lambda}=0.7$ (black line), alongside a curve representing our best-fit $H_0$ using Planck values for $\Omega_m=0.315$ and $\Omega_{\Lambda}=0.685$ (red dashed line). The negligible difference between these curves ($0.06$\% in $H_0$) demonstrates the insensitivity of our $H_0$ estimate to reasonable variations in $\Omega_m$ and $\Omega_{\Lambda}$ at these low redshifts.

We also plot the curve corresponding to \cite{Planck2020} values for $H_0$, $\Omega_m$ and $\Omega_{\Lambda}$ (blue line), which shows an offset from our best-fit model. Quantitatively, an average change of 0.0522 in $\mu_A$ over our fitting range ($z>0.023$) would be required to reconcile our measurements with the Planck $H_0$ value of 67.4 km s$^{-1}$ Mpc$^{-1}$. \cite{Scolnic2025} have measured that a distance to Coma of $111.8\pm1.8$ Mpc would be needed to align our DESI measurements with Planck. 

The lower panel of Fig. \ref{Hubble_diagram} illustrates the residuals in comparison to the best fit results for a flat $\Lambda$CDM cosmology, as detailed in equation \ref{k_modulus}. The observed larger $\mu_A$ at $z < 0.023$ can be attributed to the effect of LSS. This trend can be well explained by a single attractor model, which shows infall toward the attractor's centre and backside infall. This trend is also evident in velocity field reconstruction models. We have included three pv reconstructions (2M++ from \cite{Said2020}, 2MRS from \cite{Lilow2021} and 2M++ from \cite{Carrick2015}), which all show the same signature as the DESI FP data. The smoother curves in these models result from their use of linear theory for reconstruction and smoothing to a scale usually around 4 Mpc, but they differ from one work to the other.

\begin{figure*}
	\includegraphics[width=\textwidth]{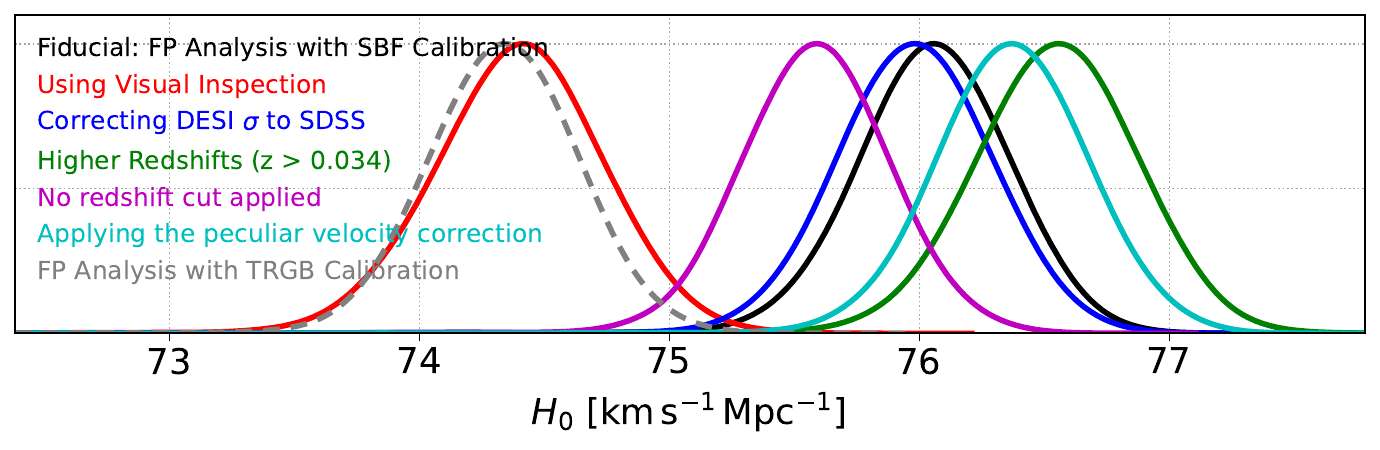}
    \caption{The posterior distribution for the Hubble constant $H_0$ derived from our fiducial DESI FP analysis represented by the black probability density. Additional probability density distributions, displayed in various colours, illustrate the impact of different sources of systematics on the $H_0$ measurement process. For the Python code and data used to reproduce this plot, see \href{https://github.com/KSaid-1/DESI_fuji_FP/tree/main/Plots/Fig.8}{this link.}}
    \label{H0_sys}
\end{figure*}

To address potential systematic errors, we conducted an analysis, revisiting our calculations while considering several sources of systematic bias. Figure \ref{H0_sys} presents the posterior distributions of the Hubble constant obtained from the DESI FP data after applying various sources of systematics to it. In Figure \ref{H0_sys}, the black probability density distribution represents our fiducial measurement. Although other potential sources of systematics may exist, the following are the most evident and notable.

\begin{enumerate}
    \item Using Visual Inspection: Despite implementing selection criteria to isolate pure elliptical galaxies during our sample selection process, we initially followed the conventional approach used in previous FP studies by performing visual inspection to classify galaxies as ellipticals or non-ellipticals. This visual classification was conducted to assess whether such subjective morphological filtering improves the FP constraints. The resulting probability density after outright removal of all identified non-ellipticals from this visual classification process is shown in red, revealing a shift in the mean value of $H_0$ of about $1.6$ $\mathrm{km \ s^{-1} Mpc^{-1}}$. However, instead of just incorporating this posterior into our systematic error budget, we performed a few statistical analyses to determine how various galaxy parameters differ between visually classified ellipticals and non-ellipticals.

    When comparing galaxies classified as ellipticals versus non-ellipticals through visual inspection, we found several statistically significant differences between the populations. We employed the Mann-Whitney U test due to its robustness against non-Gaussian distributions common in our data. To quantify effect sizes, we calculated Cohen's d values.
    
    The most striking difference appeared in effective radius, where visually classified non-ellipticals exhibited substantially larger sizes (mean=0.475) than ellipticals (mean=0.251), producing a large effect size (Cohen's d=0.861).
    
    These findings reveal that visual morphological classification introduces systematic biases toward smaller galaxies, which would significantly impact our FP analysis. Given that such subjective selection cannot be accurately modeled in selection function corrections, we opted not to apply visual classification filters to our final sample, despite observing shifts in the resulting Hubble constant estimates. Nevertheless, presenting it here serves as a point of interest for further investigation, particularly with the anticipated DESI year 1 data, which promises a substantially larger sample for a more in-depth analysis.

    \item Correcting DESI $\sigma$ to SDSS values: A key component of the FP parameters is the stellar velocity dispersion $\sigma$. An important consideration for DESI data is whether the signal-to-noise ratio is sufficient to obtain accurate measurements of velocity dispersion. To address this concern, we perform a comprehensive internal and external consistency check analysis in this paper (refer to Figs. \ref{internal_consistency} and \ref{external_consistency_v3}). When we compare DESI velocity dispersion measurements to those of SDSS, we observe a slight difference, as shown in Figure \ref{external_consistency_v3}. To assess this potential source of systematic bias, we employ linear fitting to rectify all velocity dispersion measurements across our dataset, encompassing not only the overlapping DESI and SDSS measurements but all DESI data. We then repeat the entire process of fitting the FP, perform re-calibration, and re-fit the Hubble diagram using these corrected velocity dispersion values. This results in a Hubble constant measurement of $H_0 = 75.97\pm0.34$ $\mathrm{km \ s^{-1} Mpc^{-1}}$, which we identify as the least influential source of systematic bias among the five we have identified. The posterior distribution for this measurement is illustrated in Figure \ref{H0_sys} as the blue probability density.

    \item Higher Redshift Cut ($z > 0.034$): In the process of constructing the Hubble diagram and fitting for the Hubble constant, we adopted the practice recommended by \cite{Riess2022} of implementing a redshift cut to exclude low-redshift galaxies below redshift, $z < 0.023$ to mitigate the impact of peculiar velocities. In this systematic analysis, we examine the consequences of applying a more stringent redshift cut of $z < 0.034$. This specific redshift limit was chosen to exclude any data used in the Zero-point calibration process. The resulting probability density for this cut is represented in green in Figure \ref{H0_sys}. Applying this stricter redshift cut yields a Hubble constant measurement of $H_0 = 76.54\pm0.36$ $\mathrm{km \ s^{-1} Mpc^{-1}}$, which is slightly higher than our fiducial value.

    \item No Redshift Cut Applied: In this systematic assessment, we conducted the Hubble constant measurement without implementing any redshift cut, including all the FP data, even the low-redshift galaxies. The resulting probability density is shown in purple, and it provides a Hubble constant value of $H_0 = 75.58\pm0.32$ $\mathrm{km \ s^{-1} Mpc^{-1}}$. This value is slightly lower than our fiducial Hubble constant measurement.

    \item Applying the peculiar velocity correction: Throughout our cosmological fitting, we utilized the redshift in the CMB frame, $z_{\rm cmb}$. However, peculiar velocities can introduce systematic effects. To assess this, we corrected the redshift from the CMB frame, which accounts for our own motion, to the cosmological redshift, incorporating peculiar velocities. This correction utilized the default option of the pvhub\footnote{https://github.com/KSaid-1/pvhub} velocity field maps developed by \cite{Carr2022}. This approach was identified by \cite{Peterson2022} as the optimal method for reducing Hubble residuals. Implementing this correction resulted in a slightly higher value for the Hubble constant: $H_0 = 76.36\pm0.33$ $\mathrm{km \ s^{-1} Mpc^{-1}}$. 

    \item FP Analysis with TRGB Calibration: In our primary cosmology fitting, we utilized the FP data calibrated with the absolute distance to the Coma cluster, as measured using the Surface Brightness Fluctuation method. We introduced an alternative distance calibration using the Tip of the Red Giant Branch (TRGB) method. Using the TRGB method to measure the distance to Leo I group, and based on the relative distance between Leo I and Coma cluster, \cite{Sakai1997} derived a distance modulus of $\mu = 35.03\pm 0.37$ for the Coma cluster. Comparing this absolute distance obtained via the TRGB calibration to our FP distances, we constructed a new catalogue of absolute distances for our sample. Subsequently, we created a Hubble diagram and re-measured the Hubble constant, yielding a value of $H_0 = 74.33\pm0.31$ $\mathrm{km \ s^{-1} Mpc^{-1}}$. This systematic shift represents the most substantial bias among all the potential sources of systematics we have identified. The probability density for this specific analysis is visualized in grey dotted line.
\end{enumerate}

Concluding our investigation into potential systematic biases, we proceeded to evaluate the systematic error linked to our Hubble constant value. This analysis incorporated all MCMC chains except two: the chain that removed non-ellipticals based on visual inspection and the chain that used TRGB distance calibration to the Coma cluster. The exclusion of the TRGB-calibrated chain is justified by its representation of a systematic bias intrinsic to the calibration method, rather than an issue within the FP fitting process. Moreover, the statistical error arising from the zero-point calibration is anticipated to address this aspect as well.

\begin{figure*}
\centering
\begin{subfigure}[t]{\textwidth}
\centering

\includegraphics[width=\textwidth]{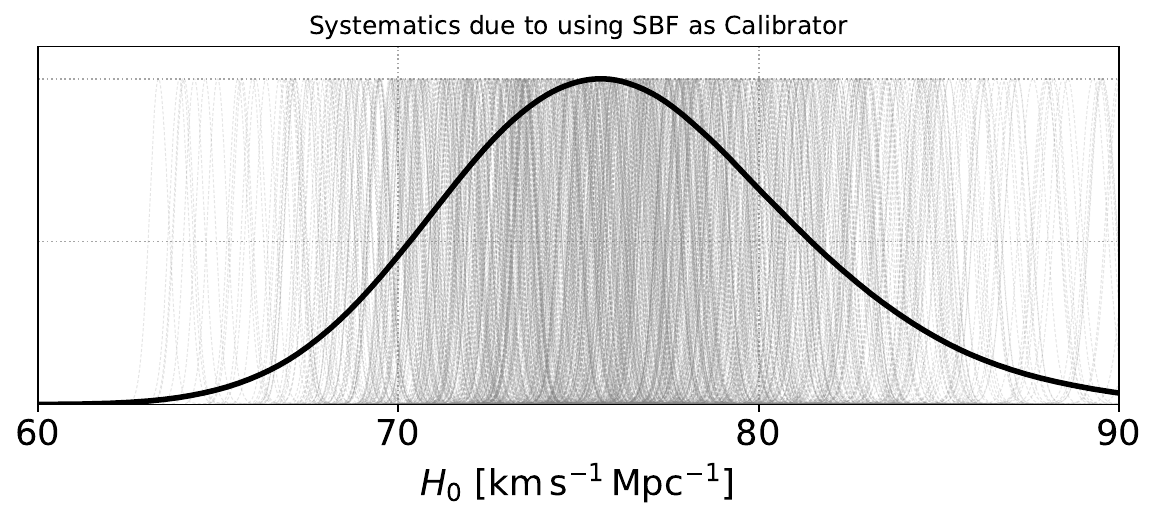}
\end{subfigure}

\begin{subfigure}[t]{\textwidth}
\vspace{-1.25cm}
\centering
\includegraphics[width=\textwidth]{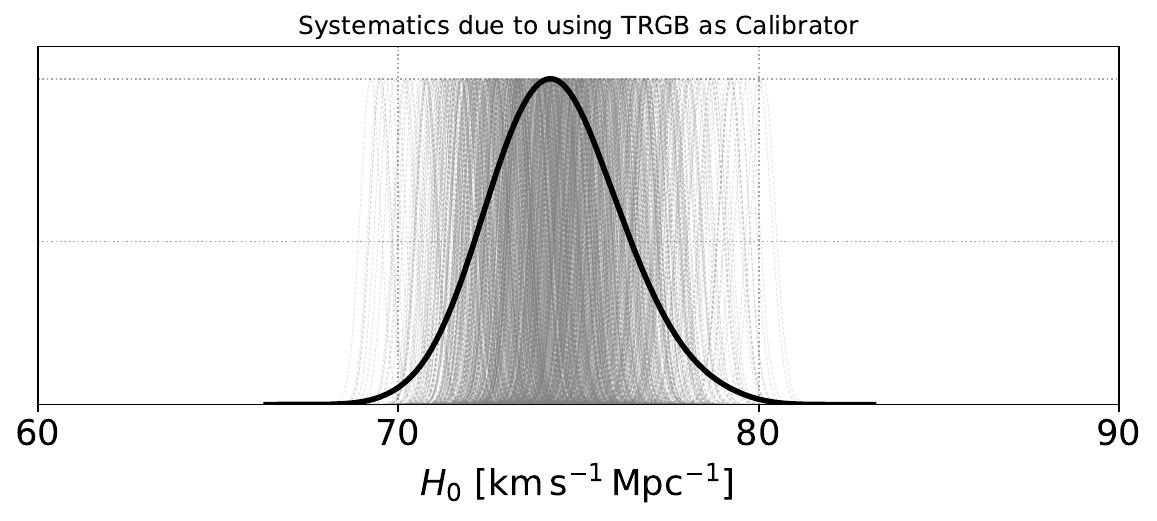}
 \end{subfigure}
    \caption{MCMC sampling of the posterior for $H_0$ to assess systematics arising from the zero-point calibration. The top panel displays 1000 MCMC chains of the $H_0$ posterior as grey dotted lines, derived using randomly sampled distances to the Coma cluster, given its measured distance and associated error from SBF. The solid line represents the combined chain used to quantify systematic errors introduced by the calibration process when using SBF. The bottom panel mirrors the top one but employs the distance to Coma cluster derived from the TRGB method. For the Python code and data used to reproduce this plot, see \href{https://github.com/KSaid-1/DESI_fuji_FP/tree/main/Plots/Fig.9}{this link.}}
    \label{H0_sys_cal}
\end{figure*}

To assess statistical uncertainties associated with the zero-point calibration process, we generated 1000 distances to the Coma cluster based on the measured distance and its associated uncertainties ($99.1\pm5.8$ Mpc.) derived from surface brightness fluctuation \citep{Jensen2021}. For each of these 1000 distances, we repeated the zero-point calibration process, generating a new catalogue used to fit the Hubble constant. The individual probability densities are depicted in the top panel of Figure \ref{H0_sys_cal} as dotted grey lines, while the combined chain is represented by the solid black line, serving to quantify the statistical  errors linked to the SBF calibration process, resulting in a value of $\pm4.86$ $\mathrm{km \ s^{-1} Mpc^{-1}}$. A similar process was conducted for the Tip of the Red Giant Branch calibration, yielding a statistical error of $\pm1.87$ $\mathrm{km \ s^{-1} Mpc^{-1}}$, as shown in the bottom panel of Figure \ref{H0_sys_cal}. Despite the lower statistical uncertainties associated with the TRGB calibration, we adopted the SBF calibration results as the main findings in this paper. This decision is based on the fact that the TRGB distance to Coma is not a direct measurement but relies on a measured distance to the Leo I group and assumes a known relative distance between Leo I and Coma cluster \citep{Sakai1997}. This finding is particularly relevant for future data releases, where direct TRGB measurements to nearby clusters will provide more precise second-rung calibrations and should be used instead of third-rung calibrations such as the SBF.

Notably, this analysis reveals that the statistical uncertainty associated with the zero-point calibration process dominates the error budget. However, this result underscores the robustness of the FP analysis. The systematic biases arising from potential sources of systematic within the FP fitting process itself are notably smaller than the statistical errors introduced by the calibration process. Therefore, after accounting for all other sources of systematics, including the inclusion of spiral galaxies, the correction of DESI velocity dispersion to SDSS values, applying both high and low redshift cuts, and statistical uncertainties due to the calibration, our final estimate for the Hubble constant is $H_0 = 76.05\pm0.35$(statistical) $\pm0.49$(systematic FP) $\pm4.86$(statistical due to calibration) $\mathrm{km \ s^{-1} Mpc^{-1}}$.

\section{Summary}
\label{S_Discussion}
The DESI peculiar velocity survey will be approximately four times larger than the combined size of all previous peculiar velocity surveys. For the science verification sample, we adopted a similar approach to the 6dFGSv and SDSS peculiar velocity surveys \citep{Magoulas2012,Howlett2022} in order to select a clean and reliable sample of elliptical galaxies. This approach involved implementing various photometric cuts, including magnitude and colour cuts. Following the implementation of the aforementioned cuts, our selection process resulted in a sample of 6698 unique galaxies. This sample size is comparable to that of the complete 6dFGSv peculiar velocity sample which had been the largest peculiar velocity survey for a decade and helped refine our understanding of the growth rate of structure \citep{Adams2017,Qin2019,Adams2020,Said2020}.

We apply only one spectroscopic cut: redshift. At this stage, we refrained from applying any H-alpha cuts, as this aspect will be investigated more extensively using data from the first year of the survey. Our final peculiar velocity sample includes 4191 elliptical galaxies.

To construct the FP, we constructed photometric parameters such as the angular effective radius and the mean surface brightness. These parameters were obtained from the Ninth Data Release (DR9) of the DESI Legacy Imaging Surveys \citep{Dey2019}. The photometric error was calculated as $\epsilon_{\rm  phot} = [(\epsilon_r)^2 + (b\epsilon_i)^2]^{1/2} = 0.0025$~dex (<1\%). This error is one order of magnitude smaller than the total photometric error observed in the SDSS peculiar velocity survey.

Velocity dispersion serves as the third component in constructing the FP. In our study, we employed the pPXF algorithm \citep{Cappellari2017} along with the Indo-U.S. Coudé Feed Spectral Library \citep{Valdes2004} to measure velocity dispersion from DESI spectra.

In order to test the internal (per tile) and external (per survey) consistency of velocity dispersion measurements and avoid any potential systematic offsets within our data, we used the pairwise relative error. We showed that the relative error distribution, both internally and externally, followed a Gaussian distribution centered at zero with a standard deviation of one. This showed that our velocity dispersion measurements are consistent and unbiased.

To ensure the reliability of our results, we assessed the relative velocity dispersion error $\frac{\delta\sigma}{\sigma}$. Our examination revealed that 75\% of our sample exhibited a relative error of less than 10\%.

In analyzing the error budget or our FP distance measurements, we find that photometric uncertainties contribute minimally (0.0025 dex or 0.6\%) to the overall error. The uncertainty from velocity dispersion measurements is more significant at 0.0265 dex (approximately 6\%), thought still modest. The dominant source remains the intrinsic scatter of the FP relation itself at 0.1 dex (23\%). When combined, these factors result in total RMS scatter in the r-direction of the FP of approximately 0.11 dex, corresponding to uncertainties of about 24\% of the measured distances. This quantification demonstrates that while spectroscopic measurements do contribute meaningfully to the error budget, improving photometric and spectroscopic precision would yield minimal benefits for distance determination, and trying to reduce the intrinsic scatter of the FP relation should be the priority for upcoming surveys. 

The fitting of a 3D Gaussian FP to our sample yielded results that were comparable to those obtained from the SDSS survey \citep{Said2020,Howlett2022} in terms of scatter in the $r$-direction of the FP. However, a significant improvement was observed when compared to the analysis based on the 6dFGSv survey.

Independent of the FP, we defined Coma cluster membership. Using this criterion, we identified 226 galaxies belonging to the Coma cluster. These galaxies were then used in our Zero-point calibration for the FP. This calibration utilized the absolute distance to the Coma cluster, which was determined by \cite{Jensen2021} through the Surface Brightness Fluctuation method.

After calibrating our sample, we proceeded to construct the Hubble diagram and estimate the Hubble constant. Our final result for the Hubble constant is $H_0 = 76.05\pm0.35$ (statistical) $\pm0.49$ (systematic FP) $\pm4.86$ (statistical due to calibration) $\mathrm{km \ s^{-1} Mpc^{-1}}$.

While our measured value of the Hubble constant, $H_0$, is within $2\sigma$ of the measurement derived from cosmic microwave background (CMB) anisotropies (e.g., \citealt{Planck2020}), it notably aligns well within $1\sigma$ with local Hubble constant determinations based on different distance indicators (e.g., \citealt{Riess2022}). This alignment includes includes the most recent findings from CosmicFlows-4 \citep{Kourkchi2020,Kourkchi2022} as discussed in \cite{Said2023}, where a comprehensive review of Hubble constant measurements from  the Tully-Fisher relation is provided.  Additionally, our measurement is consistent with the first standard siren measurement from the GW170817 analysis performed with DESI data \citep{Ballard2023}.

\section{CONCLUSIONS AND FUTURE DIRECTIONS}
\label{S_conclusions}
This paper underscores the capability of the Dark Energy Spectroscopic Instrument (DESI) to deliver reliable velocity dispersion measurements, thereby facilitating the application of the FP analysis and the subsequent constraint of critical cosmological parameters, including the Hubble constant and the growth rate of cosmic structure.

In this paper, we present an analysis of the Hubble constant ($H_0$) based on FP measurements derived from a sample of 4191 galaxies within the redshift range of 0.01 to 0.1. Systematic uncertainties are explored, including potential biases introduced by spiral galaxies, velocity dispersion calibration, and redshift cuts. The final result, $H_0 = 76.05\pm0.35$(statistical) $\pm0.49$(systematic FP) $\pm4.86$(statistical due to calibration) $\mathrm{km \ s^{-1} Mpc^{-1}}$, demonstrates an agreement with previous Hubble constant measurements from other distance indicators. 

Several avenues for expanding our analysis are on the horizon. Foremost is the utilization of the forthcoming DESI year 1 FP data (Ross et al. in preparation), expected to be substantially larger ($\sim$100k elliptical galaxies), providing not only enhanced statistical power but also greater control over systematic uncertainties, which currently represent the primary source of uncertainty, particularly in the zero-point calibration process. Currently, our calibration relies on a single cluster, Coma, and a solitary source within the cluster, NGC 4874. However, we anticipate a substantial improvement in precision with the inclusion of multiple clusters and numerous sources within each of these clusters. This advancement will become feasible with the availability of the group catalogue from the full DESI dataset.

Future research could explore the potential of using stellar population to enhance the precision and accuracy of the FP as a distance indicator. Recent work by \citep{D'Eugenio2024} discuss the concept of a "hyperplane" for early-type galaxies, which incorporates stellar population observables alongside the traditional FP parameters. 

Future work can also focus on extending our peculiar velocity survey to higher redshifts using Brightest Cluster Galaxies (BCGs). As the most luminous galaxies in the Universe, BCGs offer the potential to probe peculiar velocities out to much greater distances than possible with typical spiral and elliptical galaxies. We plan to utilize both the FP relation and the Metric Plane \citep{Lauer2014} for BCGs, with the latter offering intrinsically less scatter and thus more precise distance measurements. The DESI survey, particularly the combination of Bright Galaxy Survey (BGS) and Luminous Red Galaxy (LRG) data, is expected to provide a volume-limited sample of BCG-like galaxies. Moreover, BCGs are less affected by selection biases that impact normal ellipticals. This extension will enable us to investigate deviations from normal kinematic Hubble flow expansion out to $z\sim0.15$, potentially shedding light on the existence and effects of large-scale structures such as a large cosmic voids on cosmological parameters like $H_0$. 

Extending our distance measurements to higher redshifts will enable simultaneous fitting of cosmological parameters like the Hubble constant, deceleration, and jerk parameters. Combining our distance measurements with other DESI datasets, such as the Tully-Fisher relation, can further mitigate systematics. Additionally, incorporating our data with external datasets like SN Ia measurements holds promise. However, a key part of making these advancements is putting more effort into improving the zero-point calibration, which this paper shows is a major source of uncertainties.

Exploring other scientific domains, like the measurements of the amplitude and directions of the Bulk Flow, necessitates complete sky coverage. While DESI encompasses 14,000 square degrees in the northern hemisphere, upcoming surveys such as WALLABY \citep{Courtois2023b}, which employs the Australian SKA Pathfinder (ASKAP), and the 4MOST Hemisphere Survey \citep{Taylor2023}, are set to encompass the southern hemisphere. This expanded coverage promises numerous scientific possibilities that would be unattainable with just half of the sky. It is noteworthy that surveys like WALLABY will operate in entirely different wavelengths. Such diversity is invaluable for testing phenomena like galaxy bias across varying wavelengths, and might unravel sources of systematics that we have not thought about yet.

\section*{Acknowledgements}
 KS, CH, and TMD acknowledge support from the Australian Government through the Australian Research Council’s Laureate Fellowship funding scheme (project FL180100168) and the Australian Research Council Centre of Excellence for Gravitational Wave Discovery (OzGrav), through project number CE230100016.

KS would like to thank Francesco D'Eugenio for many useful discussions realted to pPXF and the associated stellar library templates. 

This research used data obtained with the Dark Energy Spectroscopic Instrument (DESI). DESI construction and operations is managed by the Lawrence Berkeley National Laboratory. This material is based upon work supported by the U.S. Department of Energy, Office of Science, Office of High-Energy Physics, under Contract No. DE–AC02–05CH11231, and by the National Energy Research Scientific Computing Center, a DOE Office of Science User Facility under the same contract. Additional support for DESI was provided by the U.S. National Science Foundation (NSF), Division of Astronomical Sciences under Contract No. AST-0950945 to the NSF’s National Optical-Infrared Astronomy Research Laboratory; the Science and Technology Facilities Council of the United Kingdom; the Gordon and Betty Moore Foundation; the Heising-Simons Foundation; the French Alternative Energies and Atomic Energy Commission (CEA); the National Council of Science and Technology of Mexico (CONACYT); the Ministry of Science and Innovation of Spain (MICINN), and by the DESI Member Institutions: www.desi.lbl.gov/collaborating-institutions. The DESI collaboration is honored to be permitted to conduct scientific research on Iolkam Du’ag (Kitt Peak), a mountain with particular significance to the Tohono O’odham Nation. Any opinions, findings, and conclusions or recommendations expressed in this material are those of the author(s) and do not necessarily reflect the views of the U.S. National Science Foundation, the U.S. Department of Energy, or any of the listed funding agencies.

The DESI Legacy Imaging Surveys consist of three individual and complementary projects: the Dark Energy Camera Legacy Survey (DECaLS), the Beijing-Arizona Sky Survey (BASS), and the Mayall z-band Legacy Survey (MzLS). DECaLS, BASS and MzLS together include data obtained, respectively, at the Blanco telescope, Cerro Tololo Inter-American Observatory, NSF’s NOIRLab; the Bok telescope, Steward Observatory, University of Arizona; and the Mayall telescope, Kitt Peak National Observatory, NOIRLab. NOIRLab is operated by the Association of Universities for Research in Astronomy (AURA) under a cooperative agreement with the National Science Foundation. Pipeline processing and analyses of the data were supported by NOIRLab and the Lawrence Berkeley National Laboratory. Legacy Surveys also uses data products from the Near-Earth Object Wide-field Infrared Survey Explorer (NEOWISE), a project of the Jet Propulsion Laboratory/California Institute of Technology, funded by the National Aeronautics and Space Administration. Legacy Surveys was supported by: the Director, Office of Science, Office of High Energy Physics of the U.S. Department of Energy; the National Energy Research Scientific Computing Center, a DOE Office of Science User Facility; the U.S. National Science Foundation, Division of Astronomical Sciences; the National Astronomical Observatories of China, the Chinese Academy of Sciences and the Chinese National Natural Science Foundation. LBNL is managed by the Regents of the University of California under contract to the U.S. Department of Energy. The complete acknowledgments can be found at \url{https://www.legacysurvey.org/.}
\section*{Data Availability}
All data and codes necessary to reproduce the analyses presented in this paper are publicly available. The repository containing both the final catalogue and the analysis codes can be accessed at \url{https://github.com/KSaid-1/DESI_fuji_FP} \href{https://github.com/KSaid-1/DESI_fuji_FP}. All data points shown in the figures are available in a machine-readable form on \href{https://doi.org/10.5281/zenodo.13363598}{https://doi.org/10.5281/zenodo.13363598}.

\bibliographystyle{mnras}
\bibliography{example} 

\begin{thebibliography}{}
\makeatletter
\relax
\def\mn@urlcharsother{\let\do\@makeother \do\$\do\&\do\#\do\^\do\_\do\%\do\~}
\def\mn@doi{\begingroup\mn@urlcharsother \@ifnextchar [ {\mn@doi@} {\mn@doi@[]}}
\def\mn@doi@[#1]#2{\def\@tempa{#1}\ifx\@tempa\@empty \href {http://dx.doi.org/#2} {doi:#2}\else \href {http://dx.doi.org/#2} {#1}\fi \endgroup}
\def\mn@eprint#1#2{\mn@eprint@#1:#2::\@nil}
\def\mn@eprint@arXiv#1{\href {http://arxiv.org/abs/#1} {{\tt arXiv:#1}}}
\def\mn@eprint@dblp#1{\href {http://dblp.uni-trier.de/rec/bibtex/#1.xml} {dblp:#1}}
\def\mn@eprint@#1:#2:#3:#4\@nil{\def\@tempa {#1}\def\@tempb {#2}\def\@tempc {#3}\ifx \@tempc \@empty \let \@tempc \@tempb \let \@tempb \@tempa \fi \ifx \@tempb \@empty \def\@tempb {arXiv}\fi \@ifundefined {mn@eprint@\@tempb}{\@tempb:\@tempc}{\expandafter \expandafter \csname mn@eprint@\@tempb\endcsname \expandafter{\@tempc}}}

\bibitem[\protect\citeauthoryear{{Abell}, {Corwin}  \& {Olowin}}{{Abell} et~al.}{1989}]{Abell:1989}
{Abell} G.~O.,  {Corwin} Harold~G. J.,   {Olowin} R.~P.,  1989, \mn@doi [\apjs] {10.1086/191333}, \href {https://ui.adsabs.harvard.edu/abs/1989ApJS...70....1A} {70, 1}

\bibitem[\protect\citeauthoryear{{Abolfathi} et~al.,}{{Abolfathi} et~al.}{2018}]{Abolfathi2018}
{Abolfathi} B.,  et~al., 2018, \mn@doi [\apjs] {10.3847/1538-4365/aa9e8a}, \href {https://ui.adsabs.harvard.edu/abs/2018ApJS..235...42A} {235, 42}

\bibitem[\protect\citeauthoryear{{Adams} \& {Blake}}{{Adams} \& {Blake}}{2017}]{Adams2017}
{Adams} C.,  {Blake} C.,  2017, \mn@doi [\mnras] {10.1093/mnras/stx1529}, \href {https://ui.adsabs.harvard.edu/abs/2017MNRAS.471..839A} {471, 839}

\bibitem[\protect\citeauthoryear{{Adams} \& {Blake}}{{Adams} \& {Blake}}{2020}]{Adams2020}
{Adams} C.,  {Blake} C.,  2020, \mn@doi [\mnras] {10.1093/mnras/staa845}, \href {https://ui.adsabs.harvard.edu/abs/2020MNRAS.494.3275A} {494, 3275}

\bibitem[\protect\citeauthoryear{{Ballard} et~al.,}{{Ballard} et~al.}{2023}]{Ballard2023}
{Ballard} W.,  et~al., 2023, \mn@doi [Research Notes of the American Astronomical Society] {10.3847/2515-5172/ad0eda}, \href {https://ui.adsabs.harvard.edu/abs/2023RNAAS...7..250B} {7, 250}

\bibitem[\protect\citeauthoryear{{Bell}, {Said}, {Davis}  \& {Jarrett}}{{Bell} et~al.}{2023}]{Bell2023}
{Bell} R.,  {Said} K.,  {Davis} T.,   {Jarrett} T.~H.,  2023, \mn@doi [\mnras] {10.1093/mnras/stac3407}, \href {https://ui.adsabs.harvard.edu/abs/2023MNRAS.519..102B} {519, 102}

\bibitem[\protect\citeauthoryear{{Bernardi} et~al.,}{{Bernardi} et~al.}{2003a}]{Bernardi2003b}
{Bernardi} M.,  et~al., 2003a, \mn@doi [\aj] {10.1086/374256}, \href {https://ui.adsabs.harvard.edu/abs/2003AJ....125.1849B} {125, 1849}

\bibitem[\protect\citeauthoryear{{Bernardi} et~al.,}{{Bernardi} et~al.}{2003b}]{Bernardi2003}
{Bernardi} M.,  et~al., 2003b, \mn@doi [\aj] {10.1086/367794}, \href {https://ui.adsabs.harvard.edu/abs/2003AJ....125.1866B} {125, 1866}

\bibitem[\protect\citeauthoryear{{Boubel}, {Colless}, {Said}  \& {Staveley-Smith}}{{Boubel} et~al.}{2024a}]{Boubel2024b}
{Boubel} P.,  {Colless} M.,  {Said} K.,   {Staveley-Smith} L.,  2024a, \mn@doi [arXiv e-prints] {10.48550/arXiv.2408.03660}, \href {https://ui.adsabs.harvard.edu/abs/2024arXiv240803660B} {p. arXiv:2408.03660}

\bibitem[\protect\citeauthoryear{{Boubel}, {Colless}, {Said}  \& {Staveley-Smith}}{{Boubel} et~al.}{2024b}]{Boubel2024a}
{Boubel} P.,  {Colless} M.,  {Said} K.,   {Staveley-Smith} L.,  2024b, \mn@doi [\mnras] {10.1093/mnras/stae1122}, \href {https://ui.adsabs.harvard.edu/abs/2024MNRAS.531...84B} {531, 84}

\bibitem[\protect\citeauthoryear{{Campbell} et~al.,}{{Campbell} et~al.}{2014}]{Campbell2014}
{Campbell} L.~A.,  et~al., 2014, \mn@doi [\mnras] {10.1093/mnras/stu1198}, \href {https://ui.adsabs.harvard.edu/abs/2014MNRAS.443.1231C} {443, 1231}

\bibitem[\protect\citeauthoryear{{Cappellari}}{{Cappellari}}{2017}]{Cappellari2017}
{Cappellari} M.,  2017, \mn@doi [\mnras] {10.1093/mnras/stw3020}, \href {https://ui.adsabs.harvard.edu/abs/2017MNRAS.466..798C} {466, 798}

\bibitem[\protect\citeauthoryear{{Cappellari}}{{Cappellari}}{2022}]{Cappellari2022}
{Cappellari} M.,  2022, \mn@doi [arXiv e-prints] {10.48550/arXiv.2208.14974}, \href {https://ui.adsabs.harvard.edu/abs/2022arXiv220814974C} {p. arXiv:2208.14974}

\bibitem[\protect\citeauthoryear{{Cappellari} \& {Emsellem}}{{Cappellari} \& {Emsellem}}{2004}]{Cappellari2004}
{Cappellari} M.,  {Emsellem} E.,  2004, \mn@doi [\pasp] {10.1086/381875}, \href {https://ui.adsabs.harvard.edu/abs/2004PASP..116..138C} {116, 138}

\bibitem[\protect\citeauthoryear{{Carr}, {Davis}, {Scolnic}, {Said}, {Brout}, {Peterson}  \& {Kessler}}{{Carr} et~al.}{2022}]{Carr2022}
{Carr} A.,  {Davis} T.~M.,  {Scolnic} D.,  {Said} K.,  {Brout} D.,  {Peterson} E.~R.,   {Kessler} R.,  2022, \mn@doi [\pasa] {10.1017/pasa.2022.41}, \href {https://ui.adsabs.harvard.edu/abs/2022PASA...39...46C} {39, e046}

\bibitem[\protect\citeauthoryear{{Carrick}, {Turnbull}, {Lavaux}  \& {Hudson}}{{Carrick} et~al.}{2015}]{Carrick2015}
{Carrick} J.,  {Turnbull} S.~J.,  {Lavaux} G.,   {Hudson} M.~J.,  2015, \mn@doi [\mnras] {10.1093/mnras/stv547}, \href {https://ui.adsabs.harvard.edu/abs/2015MNRAS.450..317C} {450, 317}

\bibitem[\protect\citeauthoryear{{Carroll}}{{Carroll}}{2001}]{Carroll2001}
{Carroll} S.~M.,  2001, \mn@doi [Living Reviews in Relativity] {10.12942/lrr-2001-1}, \href {http://adsabs.harvard.edu/abs/2001LRR.....4....1C} {4, 1}

\bibitem[\protect\citeauthoryear{{Chambers} et~al.,}{{Chambers} et~al.}{2016}]{Chambers2016}
{Chambers} K.~C.,  et~al., 2016, \mn@doi [arXiv e-prints] {10.48550/arXiv.1612.05560}, \href {https://ui.adsabs.harvard.edu/abs/2016arXiv161205560C} {p. arXiv:1612.05560}

\bibitem[\protect\citeauthoryear{{Chilingarian}, {Melchior}  \& {Zolotukhin}}{{Chilingarian} et~al.}{2010}]{Chilingarian2010}
{Chilingarian} I.~V.,  {Melchior} A.-L.,   {Zolotukhin} I.~Y.,  2010, \mn@doi [\mnras] {10.1111/j.1365-2966.2010.16506.x}, \href {https://ui.adsabs.harvard.edu/abs/2010MNRAS.405.1409C} {405, 1409}

\bibitem[\protect\citeauthoryear{{Ciotti}, {Lanzoni}  \& {Renzini}}{{Ciotti} et~al.}{1996}]{Ciotti1996}
{Ciotti} L.,  {Lanzoni} B.,   {Renzini} A.,  1996, \mn@doi [\mnras] {10.1093/mnras/282.1.1}, \href {https://ui.adsabs.harvard.edu/abs/1996MNRAS.282....1C} {282, 1}

\bibitem[\protect\citeauthoryear{{Colless}, {Saglia}, {Burstein}, {Davies}, {McMahan}  \& {Wegner}}{{Colless} et~al.}{2001}]{Colless2001}
{Colless} M.,  {Saglia} R.~P.,  {Burstein} D.,  {Davies} R.~L.,  {McMahan} R.~K.,   {Wegner} G.,  2001, \mn@doi [\mnras] {10.1046/j.1365-8711.2001.04044.x}, \href {https://ui.adsabs.harvard.edu/abs/2001MNRAS.321..277C} {321, 277}

\bibitem[\protect\citeauthoryear{{Courtois} et~al.,}{{Courtois} et~al.}{2023a}]{Courtois2023b}
{Courtois} H.~M.,  et~al., 2023a, \mn@doi [\mnras] {10.1093/mnras/stac3246}, \href {https://ui.adsabs.harvard.edu/abs/2023MNRAS.519.4589C} {519, 4589}

\bibitem[\protect\citeauthoryear{{Courtois}, {Dupuy}, {Guinet}, {Baulieu}, {Ruppin}  \& {Brenas}}{{Courtois} et~al.}{2023b}]{Courtois2023}
{Courtois} H.~M.,  {Dupuy} A.,  {Guinet} D.,  {Baulieu} G.,  {Ruppin} F.,   {Brenas} P.,  2023b, \mn@doi [\aap] {10.1051/0004-6361/202245331}, \href {https://ui.adsabs.harvard.edu/abs/2023A&A...670L..15C} {670, L15}

\bibitem[\protect\citeauthoryear{{DESI Collaboration} et~al.,}{{DESI Collaboration} et~al.}{2016a}]{DESI_Collaboration2016}
{DESI Collaboration} et~al., 2016a, \mn@doi [arXiv e-prints] {10.48550/arXiv.1611.00036}, \href {https://ui.adsabs.harvard.edu/abs/2016arXiv161100036D} {p. arXiv:1611.00036}

\bibitem[\protect\citeauthoryear{{DESI Collaboration} et~al.,}{{DESI Collaboration} et~al.}{2016b}]{DESICollaboration2016b}
{DESI Collaboration} et~al., 2016b, \mn@doi [arXiv e-prints] {10.48550/arXiv.1611.00037}, \href {https://ui.adsabs.harvard.edu/abs/2016arXiv161100037D} {p. arXiv:1611.00037}

\bibitem[\protect\citeauthoryear{{DESI Collaboration} et~al.,}{{DESI Collaboration} et~al.}{2022}]{DESI_Collaboration2022}
{DESI Collaboration} et~al., 2022, \mn@doi [\aj] {10.3847/1538-3881/ac882b}, \href {https://ui.adsabs.harvard.edu/abs/2022AJ....164..207D} {164, 207}

\bibitem[\protect\citeauthoryear{{DESI Collaboration} et~al.,}{{DESI Collaboration} et~al.}{2023a}]{DESI_Collaboration2023b}
{DESI Collaboration} et~al., 2023a, \mn@doi [arXiv e-prints] {10.48550/arXiv.2306.06307}, \href {https://ui.adsabs.harvard.edu/abs/2023arXiv230606307D} {p. arXiv:2306.06307}

\bibitem[\protect\citeauthoryear{{DESI Collaboration} et~al.,}{{DESI Collaboration} et~al.}{2023b}]{DESI_Collaboration2023}
{DESI Collaboration} et~al., 2023b, \mn@doi [arXiv e-prints] {10.48550/arXiv.2306.06308}, \href {https://ui.adsabs.harvard.edu/abs/2023arXiv230606308D} {p. arXiv:2306.06308}

\bibitem[\protect\citeauthoryear{{D'Eugenio} et~al.,}{{D'Eugenio} et~al.}{2021}]{D'Eugenio2021}
{D'Eugenio} F.,  et~al., 2021, \mn@doi [\mnras] {10.1093/mnras/stab1146}, \href {https://ui.adsabs.harvard.edu/abs/2021MNRAS.504.5098D} {504, 5098}

\bibitem[\protect\citeauthoryear{{D'Eugenio} et~al.,}{{D'Eugenio} et~al.}{2024}]{D'Eugenio2024}
{D'Eugenio} F.,  et~al., 2024, \mn@doi [\mnras] {10.1093/mnras/stae1582}, \href {https://ui.adsabs.harvard.edu/abs/2024MNRAS.532.1775D} {532, 1775}

\bibitem[\protect\citeauthoryear{{Davis}, {Hinton}, {Howlett}  \& {Calcino}}{{Davis} et~al.}{2019}]{Davis2019}
{Davis} T.~M.,  {Hinton} S.~R.,  {Howlett} C.,   {Calcino} J.,  2019, \mn@doi [\mnras] {10.1093/mnras/stz2652}, \href {https://ui.adsabs.harvard.edu/abs/2019MNRAS.490.2948D} {490, 2948}

\bibitem[\protect\citeauthoryear{{Dey} et~al.,}{{Dey} et~al.}{2019}]{Dey2019}
{Dey} A.,  et~al., 2019, \mn@doi [\aj] {10.3847/1538-3881/ab089d}, \href {https://ui.adsabs.harvard.edu/abs/2019AJ....157..168D} {157, 168}

\bibitem[\protect\citeauthoryear{{Dicus}, {Kolb}  \& {Teplitz}}{{Dicus} et~al.}{1977}]{Dicus1977}
{Dicus} D.~A.,  {Kolb} E.~W.,   {Teplitz} V.~L.,  1977, \mn@doi [\prl] {10.1103/PhysRevLett.39.168}, \href {https://ui.adsabs.harvard.edu/abs/1977PhRvL..39..168D} {39, 168}

\bibitem[\protect\citeauthoryear{{Djorgovski} \& {Davis}}{{Djorgovski} \& {Davis}}{1987}]{Djorgovski1987}
{Djorgovski} S.,  {Davis} M.,  1987, \mn@doi [\apj] {10.1086/164948}, \href {http://adsabs.harvard.edu/abs/1987ApJ...313...59D} {313, 59}

\bibitem[\protect\citeauthoryear{{Dressler}, {Faber}, {Burstein}, {Davies}, {Lynden-Bell}, {Terlevich}  \& {Wegner}}{{Dressler} et~al.}{1987}]{Dressler1987}
{Dressler} A.,  {Faber} S.~M.,  {Burstein} D.,  {Davies} R.~L.,  {Lynden-Bell} D.,  {Terlevich} R.~J.,   {Wegner} G.,  1987, \mn@doi [\apjl] {10.1086/184827}, \href {http://adsabs.harvard.edu/abs/1987ApJ...313L..37D} {313, L37}

\bibitem[\protect\citeauthoryear{{Duarte} \& {Mamon}}{{Duarte} \& {Mamon}}{2014}]{Duarte:2014}
{Duarte} M.,  {Mamon} G.~A.,  2014, \mn@doi [\mnras] {10.1093/mnras/stu378}, \href {http://cdsads.u-strasbg.fr/abs/2014MNRAS.440.1763D} {440, 1763}

\bibitem[\protect\citeauthoryear{{Duarte} \& {Mamon}}{{Duarte} \& {Mamon}}{2015}]{Duarte:2015}
{Duarte} M.,  {Mamon} G.~A.,  2015, \mn@doi [\mnras] {10.1093/mnras/stv1799}, \href {http://cdsads.u-strasbg.fr/abs/2015MNRAS.453.3848D} {453, 3848}

\bibitem[\protect\citeauthoryear{{Eke} et~al.,}{{Eke} et~al.}{2004}]{Eke:2004a}
{Eke} V.~R.,  et~al., 2004, \mn@doi [\mnras] {10.1111/j.1365-2966.2004.07408.x}, \href {http://adsabs.harvard.edu/abs/2004MNRAS.348..866E} {348, 866}

\bibitem[\protect\citeauthoryear{{Faber} \& {Jackson}}{{Faber} \& {Jackson}}{1976}]{Faber1976}
{Faber} S.~M.,  {Jackson} R.~E.,  1976, \mn@doi [\apj] {10.1086/154215}, \href {https://ui.adsabs.harvard.edu/abs/1976ApJ...204..668F} {204, 668}

\bibitem[\protect\citeauthoryear{{Faber}, {Dressler}, {Davies}, {Burstein}, {Lynden Bell}, {Terlevich}  \& {Wegner}}{{Faber} et~al.}{1987}]{Faber1987}
{Faber} S.~M.,  {Dressler} A.,  {Davies} R.~L.,  {Burstein} D.,  {Lynden Bell} D.,  {Terlevich} R.,   {Wegner} G.,  1987, in {Faber} S.~M.,  ed., Nearly Normal Galaxies. From the Planck Time to the Present. p.~175

\bibitem[\protect\citeauthoryear{{Flaugher} et~al.,}{{Flaugher} et~al.}{2015}]{Flaugher2015}
{Flaugher} B.,  et~al., 2015, \mn@doi [\aj] {10.1088/0004-6256/150/5/150}, \href {https://ui.adsabs.harvard.edu/abs/2015AJ....150..150F} {150, 150}

\bibitem[\protect\citeauthoryear{{Giani}, {Howlett}, {Said}, {Davis}  \& {Vagnozzi}}{{Giani} et~al.}{2024}]{Giani2024}
{Giani} L.,  {Howlett} C.,  {Said} K.,  {Davis} T.,   {Vagnozzi} S.,  2024, \mn@doi [\jcap] {10.1088/1475-7516/2024/01/071}, \href {https://ui.adsabs.harvard.edu/abs/2024JCAP...01..071G} {2024, 071}

\bibitem[\protect\citeauthoryear{{Guy} et~al.,}{{Guy} et~al.}{2023}]{Guy2023}
{Guy} J.,  et~al., 2023, \mn@doi [\aj] {10.3847/1538-3881/acb212}, \href {https://ui.adsabs.harvard.edu/abs/2023AJ....165..144G} {165, 144}

\bibitem[\protect\citeauthoryear{{Hahn} et~al.,}{{Hahn} et~al.}{2023}]{Hahn2023}
{Hahn} C.,  et~al., 2023, \mn@doi [\aj] {10.3847/1538-3881/accff8}, \href {https://ui.adsabs.harvard.edu/abs/2023AJ....165..253H} {165, 253}

\bibitem[\protect\citeauthoryear{{Hildebrandt} et~al.,}{{Hildebrandt} et~al.}{2020}]{Hildebrandt2020}
{Hildebrandt} H.,  et~al., 2020, \mn@doi [\aap] {10.1051/0004-6361/201834878}, \href {https://ui.adsabs.harvard.edu/abs/2020A&A...633A..69H} {633, A69}

\bibitem[\protect\citeauthoryear{{Hoffman}, {Valade}, {Libeskind}, {Sorce}, {Tully}, {Pfeifer}, {Gottl{\"o}ber}  \& {Pomar{\`e}de}}{{Hoffman} et~al.}{2024}]{Hoffman2024}
{Hoffman} Y.,  {Valade} A.,  {Libeskind} N.~I.,  {Sorce} J.~G.,  {Tully} R.~B.,  {Pfeifer} S.,  {Gottl{\"o}ber} S.,   {Pomar{\`e}de} D.,  2024, \mn@doi [\mnras] {10.1093/mnras/stad3433}, \href {https://ui.adsabs.harvard.edu/abs/2024MNRAS.527.3788H} {527, 3788}

\bibitem[\protect\citeauthoryear{{Howlett}, {Said}, {Lucey}, {Colless}, {Qin}, {Lai}, {Tully}  \& {Davis}}{{Howlett} et~al.}{2022}]{Howlett2022}
{Howlett} C.,  {Said} K.,  {Lucey} J.~R.,  {Colless} M.,  {Qin} F.,  {Lai} Y.,  {Tully} R.~B.,   {Davis} T.~M.,  2022, \mn@doi [\mnras] {10.1093/mnras/stac1681}, \href {https://ui.adsabs.harvard.edu/abs/2022MNRAS.515..953H} {515, 953}

\bibitem[\protect\citeauthoryear{{Hudson}, {Lucey}, {Smith}  \& {Steel}}{{Hudson} et~al.}{1997}]{Hudson1997}
{Hudson} M.~J.,  {Lucey} J.~R.,  {Smith} R.~J.,   {Steel} J.,  1997, \mn@doi [\mnras] {10.1093/mnras/291.3.488}, \href {https://ui.adsabs.harvard.edu/abs/1997MNRAS.291..488H} {291, 488}

\bibitem[\protect\citeauthoryear{{Hut}}{{Hut}}{1977}]{Hut1977}
{Hut} P.,  1977, \mn@doi [Physics Letters B] {10.1016/0370-2693(77)90139-3}, \href {https://ui.adsabs.harvard.edu/abs/1977PhLB...69...85H} {69, 85}

\bibitem[\protect\citeauthoryear{{Jaff{\'e}}, {Smith}, {Candlish}, {Poggianti}, {Sheen}  \& {Verheijen}}{{Jaff{\'e}} et~al.}{2015}]{Jaffe:2015}
{Jaff{\'e}} Y.~L.,  {Smith} R.,  {Candlish} G.~N.,  {Poggianti} B.~M.,  {Sheen} Y.-K.,   {Verheijen} M. A.~W.,  2015, \mn@doi [\mnras] {10.1093/mnras/stv100}, \href {https://ui.adsabs.harvard.edu/abs/2015MNRAS.448.1715J} {448, 1715}

\bibitem[\protect\citeauthoryear{{Jensen} et~al.,}{{Jensen} et~al.}{2021}]{Jensen2021}
{Jensen} J.~B.,  et~al., 2021, \mn@doi [\apjs] {10.3847/1538-4365/ac01e7}, \href {https://ui.adsabs.harvard.edu/abs/2021ApJS..255...21J} {255, 21}

\bibitem[\protect\citeauthoryear{{Jones} et~al.,}{{Jones} et~al.}{2009}]{Jones2009}
{Jones} D.~H.,  et~al., 2009, \mn@doi [\mnras] {10.1111/j.1365-2966.2009.15338.x}, \href {https://ui.adsabs.harvard.edu/abs/2009MNRAS.399..683J} {399, 683}

\bibitem[\protect\citeauthoryear{{Jorgensen}, {Franx}  \& {Kjaergaard}}{{Jorgensen} et~al.}{1995}]{Jorgensen1995}
{Jorgensen} I.,  {Franx} M.,   {Kjaergaard} P.,  1995, \mn@doi [\mnras] {10.1093/mnras/276.4.1341}, \href {https://ui.adsabs.harvard.edu/abs/1995MNRAS.276.1341J} {276, 1341}

\bibitem[\protect\citeauthoryear{{Korkidis}, {Pavlidou}, {Tassis}, {Ntormousi}, {Tomaras}  \& {Kovlakas}}{{Korkidis} et~al.}{2020}]{Korkidis:2020}
{Korkidis} G.,  {Pavlidou} V.,  {Tassis} K.,  {Ntormousi} E.,  {Tomaras} T.~N.,   {Kovlakas} K.,  2020, \mn@doi [\aap] {10.1051/0004-6361/201937337}, \href {https://ui.adsabs.harvard.edu/abs/2020A&A...639A.122K} {639, A122}

\bibitem[\protect\citeauthoryear{{Kormendy}}{{Kormendy}}{1977}]{Kormendy1977}
{Kormendy} J.,  1977, \mn@doi [\apj] {10.1086/155687}, \href {https://ui.adsabs.harvard.edu/abs/1977ApJ...218..333K} {218, 333}

\bibitem[\protect\citeauthoryear{{Kourkchi} et~al.,}{{Kourkchi} et~al.}{2020}]{Kourkchi2020}
{Kourkchi} E.,  et~al., 2020, \mn@doi [\apj] {10.3847/1538-4357/abb66b}, \href {https://ui.adsabs.harvard.edu/abs/2020ApJ...902..145K} {902, 145}

\bibitem[\protect\citeauthoryear{{Kourkchi}, {Tully}, {Courtois}, {Dupuy}  \& {Guinet}}{{Kourkchi} et~al.}{2022}]{Kourkchi2022}
{Kourkchi} E.,  {Tully} R.~B.,  {Courtois} H.~M.,  {Dupuy} A.,   {Guinet} D.,  2022, \mn@doi [\mnras] {10.1093/mnras/stac303}, \href {https://ui.adsabs.harvard.edu/abs/2022MNRAS.511.6160K} {511, 6160}

\bibitem[\protect\citeauthoryear{{Lahav} \& {Silk}}{{Lahav} \& {Silk}}{2021}]{Lahav2021}
{Lahav} O.,  {Silk} J.,  2021, \mn@doi [Nature Astronomy] {10.1038/s41550-021-01460-7}, \href {https://ui.adsabs.harvard.edu/abs/2021NatAs...5..855L} {5, 855}

\bibitem[\protect\citeauthoryear{{Lauer}, {Postman}, {Strauss}, {Graves}  \& {Chisari}}{{Lauer} et~al.}{2014}]{Lauer2014}
{Lauer} T.~R.,  {Postman} M.,  {Strauss} M.~A.,  {Graves} G.~J.,   {Chisari} N.~E.,  2014, \mn@doi [\apj] {10.1088/0004-637X/797/2/82}, \href {https://ui.adsabs.harvard.edu/abs/2014ApJ...797...82L} {797, 82}

\bibitem[\protect\citeauthoryear{{Lee} \& {Weinberg}}{{Lee} \& {Weinberg}}{1977}]{Lee1977}
{Lee} B.~W.,  {Weinberg} S.,  1977, \mn@doi [\prl] {10.1103/PhysRevLett.39.165}, \href {https://ui.adsabs.harvard.edu/abs/1977PhRvL..39..165L} {39, 165}

\bibitem[\protect\citeauthoryear{{Levi} et~al.,}{{Levi} et~al.}{2013}]{Levi2013}
{Levi} M.,  et~al., 2013, \mn@doi [arXiv e-prints] {10.48550/arXiv.1308.0847}, \href {https://ui.adsabs.harvard.edu/abs/2013arXiv1308.0847L} {p. arXiv:1308.0847}

\bibitem[\protect\citeauthoryear{{Lilow} \& {Nusser}}{{Lilow} \& {Nusser}}{2021}]{Lilow2021}
{Lilow} R.,  {Nusser} A.,  2021, \mn@doi [\mnras] {10.1093/mnras/stab2009}, \href {https://ui.adsabs.harvard.edu/abs/2021MNRAS.507.1557L} {507, 1557}

\bibitem[\protect\citeauthoryear{{Lintott} et~al.,}{{Lintott} et~al.}{2011}]{Lintott2011}
{Lintott} C.,  et~al., 2011, \mn@doi [\mnras] {10.1111/j.1365-2966.2010.17432.x}, \href {https://ui.adsabs.harvard.edu/abs/2011MNRAS.410..166L} {410, 166}

\bibitem[\protect\citeauthoryear{{Lynden-Bell}, {Faber}, {Burstein}, {Davies}, {Dressler}, {Terlevich}  \& {Wegner}}{{Lynden-Bell} et~al.}{1988}]{Lynden-Bell1988}
{Lynden-Bell} D.,  {Faber} S.~M.,  {Burstein} D.,  {Davies} R.~L.,  {Dressler} A.,  {Terlevich} R.~J.,   {Wegner} G.,  1988, \mn@doi [\apj] {10.1086/166066}, \href {https://ui.adsabs.harvard.edu/abs/1988ApJ...326...19L} {326, 19}

\bibitem[\protect\citeauthoryear{{Magoulas} et~al.,}{{Magoulas} et~al.}{2012}]{Magoulas2012}
{Magoulas} C.,  et~al., 2012, \mn@doi [\mnras] {10.1111/j.1365-2966.2012.21421.x}, \href {https://ui.adsabs.harvard.edu/abs/2012MNRAS.427..245M} {427, 245}

\bibitem[\protect\citeauthoryear{{Masters}, {Springob}  \& {Huchra}}{{Masters} et~al.}{2008}]{Masters2008}
{Masters} K.~L.,  {Springob} C.~M.,   {Huchra} J.~P.,  2008, \mn@doi [\aj] {10.1088/0004-6256/135/5/1738}, \href {https://ui.adsabs.harvard.edu/abs/2008AJ....135.1738M} {135, 1738}

\bibitem[\protect\citeauthoryear{{Miller} et~al.,}{{Miller} et~al.}{2023}]{Miller2023}
{Miller} T.~N.,  et~al., 2023, \mn@doi [arXiv e-prints] {10.48550/arXiv.2306.06310}, \href {https://ui.adsabs.harvard.edu/abs/2023arXiv230606310M} {p. arXiv:2306.06310}

\bibitem[\protect\citeauthoryear{{Minkowski}}{{Minkowski}}{1962}]{Minkowski1962}
{Minkowski} R.,  1962, in {McVittie} G.~C.,  ed., ~ Vol. 15, Problems of Extra-Galactic Research. p.~112

\bibitem[\protect\citeauthoryear{{Moster}, {Somerville}, {Maulbetsch}, {van den Bosch}, {Macci{\`o}}, {Naab}  \& {Oser}}{{Moster} et~al.}{2010}]{Moster2010}
{Moster} B.~P.,  {Somerville} R.~S.,  {Maulbetsch} C.,  {van den Bosch} F.~C.,  {Macci{\`o}} A.~V.,  {Naab} T.,   {Oser} L.,  2010, \mn@doi [\apj] {10.1088/0004-637X/710/2/903}, \href {https://ui.adsabs.harvard.edu/abs/2010ApJ...710..903M} {710, 903}

\bibitem[\protect\citeauthoryear{{Moustakas} et~al.,}{{Moustakas} et~al.}{2023}]{Moustakas2023}
{Moustakas} J.,  et~al., 2023, \mn@doi [arXiv e-prints] {10.48550/arXiv.2307.04888}, \href {https://ui.adsabs.harvard.edu/abs/2023arXiv230704888M} {p. arXiv:2307.04888}

\bibitem[\protect\citeauthoryear{{Myers} et~al.,}{{Myers} et~al.}{2023}]{Myers2023}
{Myers} A.~D.,  et~al., 2023, \mn@doi [\aj] {10.3847/1538-3881/aca5f9}, \href {https://ui.adsabs.harvard.edu/abs/2023AJ....165...50M} {165, 50}

\bibitem[\protect\citeauthoryear{{Nguyen}, {Huterer}  \& {Wen}}{{Nguyen} et~al.}{2023}]{Nguyen2023}
{Nguyen} N.-M.,  {Huterer} D.,   {Wen} Y.,  2023, \mn@doi [arXiv e-prints] {10.48550/arXiv.2302.01331}, \href {https://ui.adsabs.harvard.edu/abs/2023arXiv230201331N} {p. arXiv:2302.01331}

\bibitem[\protect\citeauthoryear{{Peebles}}{{Peebles}}{2021}]{Peebles2021}
{Peebles} P.~J.~E.,  2021, \mn@doi [arXiv e-prints] {10.48550/arXiv.2106.02672}, \href {https://ui.adsabs.harvard.edu/abs/2021arXiv210602672P} {p. arXiv:2106.02672}

\bibitem[\protect\citeauthoryear{{Peebles} \& {Ratra}}{{Peebles} \& {Ratra}}{2003}]{Peebles2003}
{Peebles} P.~J.,  {Ratra} B.,  2003, \mn@doi [Reviews of Modern Physics] {10.1103/RevModPhys.75.559}, \href {http://adsabs.harvard.edu/abs/2003RvMP...75..559P} {75, 559}

\bibitem[\protect\citeauthoryear{{Peterson} et~al.,}{{Peterson} et~al.}{2022}]{Peterson2022}
{Peterson} E.~R.,  et~al., 2022, \mn@doi [\apj] {10.3847/1538-4357/ac4698}, \href {https://ui.adsabs.harvard.edu/abs/2022ApJ...938..112P} {938, 112}

\bibitem[\protect\citeauthoryear{{Planck Collaboration} et~al.,}{{Planck Collaboration} et~al.}{2020}]{Planck2020}
{Planck Collaboration} et~al., 2020, \mn@doi [\aap] {10.1051/0004-6361/201833910}, \href {https://ui.adsabs.harvard.edu/abs/2020A&A...641A...6P} {641, A6}

\bibitem[\protect\citeauthoryear{{Press} \& {Davis}}{{Press} \& {Davis}}{1982}]{Press:1982}
{Press} W.~H.,  {Davis} M.,  1982, \mn@doi [\apj] {10.1086/160183}, \href {http://adsabs.harvard.edu/abs/1982ApJ...259..449P} {259, 449}

\bibitem[\protect\citeauthoryear{{Qin}, {Howlett}  \& {Staveley-Smith}}{{Qin} et~al.}{2019}]{Qin2019}
{Qin} F.,  {Howlett} C.,   {Staveley-Smith} L.,  2019, \mn@doi [\mnras] {10.1093/mnras/stz1576}, \href {https://ui.adsabs.harvard.edu/abs/2019MNRAS.487.5235Q} {487, 5235}

\bibitem[\protect\citeauthoryear{{Riess} et~al.,}{{Riess} et~al.}{2022}]{Riess2022}
{Riess} A.~G.,  et~al., 2022, \mn@doi [\apjl] {10.3847/2041-8213/ac5c5b}, \href {https://ui.adsabs.harvard.edu/abs/2022ApJ...934L...7R} {934, L7}

\bibitem[\protect\citeauthoryear{{Robotham} et~al.,}{{Robotham} et~al.}{2011}]{Robotham:2011}
{Robotham} A.~S.~G.,  et~al., 2011, \mn@doi [\mnras] {10.1111/j.1365-2966.2011.19217.x}, \href {http://adsabs.harvard.edu/abs/2011MNRAS.416.2640R} {416, 2640}

\bibitem[\protect\citeauthoryear{{Ruiz-Macias} et~al.,}{{Ruiz-Macias} et~al.}{2020}]{Ruiz-Macias2020}
{Ruiz-Macias} O.,  et~al., 2020, \mn@doi [Research Notes of the American Astronomical Society] {10.3847/2515-5172/abc25a}, \href {https://ui.adsabs.harvard.edu/abs/2020RNAAS...4..187R} {4, 187}

\bibitem[\protect\citeauthoryear{{Saglia}, {Colless}, {Burstein}, {Davies}, {McMahan}  \& {Wegner}}{{Saglia} et~al.}{2001}]{Saglia2001}
{Saglia} R.~P.,  {Colless} M.,  {Burstein} D.,  {Davies} R.~L.,  {McMahan} R.~K.,   {Wegner} G.,  2001, \mn@doi [\mnras] {10.1046/j.1365-8711.2001.04317.x}, \href {https://ui.adsabs.harvard.edu/abs/2001MNRAS.324..389S} {324, 389}

\bibitem[\protect\citeauthoryear{{Said}}{{Said}}{2023}]{Said2023}
{Said} K.,  2023, \mn@doi [arXiv e-prints] {10.48550/arXiv.2310.16053}, \href {https://ui.adsabs.harvard.edu/abs/2023arXiv231016053S} {p. arXiv:2310.16053}

\bibitem[\protect\citeauthoryear{{Said}, {Colless}, {Magoulas}, {Lucey}  \& {Hudson}}{{Said} et~al.}{2020}]{Said2020}
{Said} K.,  {Colless} M.,  {Magoulas} C.,  {Lucey} J.~R.,   {Hudson} M.~J.,  2020, \mn@doi [\mnras] {10.1093/mnras/staa2032}, \href {https://ui.adsabs.harvard.edu/abs/2020MNRAS.497.1275S} {497, 1275}

\bibitem[\protect\citeauthoryear{{Sakai}, {Madore}, {Freedman}, {Lauer}, {Ajhar}  \& {Baum}}{{Sakai} et~al.}{1997}]{Sakai1997}
{Sakai} S.,  {Madore} B.~F.,  {Freedman} W.~L.,  {Lauer} T.~R.,  {Ajhar} E.~A.,   {Baum} W.~A.,  1997, \mn@doi [\apj] {10.1086/303768}, \href {https://ui.adsabs.harvard.edu/abs/1997ApJ...478...49S} {478, 49}

\bibitem[\protect\citeauthoryear{{S{\'a}nchez-Bl{\'a}zquez} et~al.,}{{S{\'a}nchez-Bl{\'a}zquez} et~al.}{2006}]{Sanchez-Blazquez2006}
{S{\'a}nchez-Bl{\'a}zquez} P.,  et~al., 2006, \mn@doi [\mnras] {10.1111/j.1365-2966.2006.10699.x}, \href {https://ui.adsabs.harvard.edu/abs/2006MNRAS.371..703S} {371, 703}

\bibitem[\protect\citeauthoryear{{Sato} \& {Kobayashi}}{{Sato} \& {Kobayashi}}{1977}]{Sato1977}
{Sato} K.,  {Kobayashi} M.,  1977, \mn@doi [Progress of Theoretical Physics] {10.1143/PTP.58.1775}, \href {https://ui.adsabs.harvard.edu/abs/1977PThPh..58.1775S} {58, 1775}

\bibitem[\protect\citeauthoryear{{Saulder}, {Mieske}, {Zeilinger}  \& {Chilingarian}}{{Saulder} et~al.}{2013}]{Saulder2013}
{Saulder} C.,  {Mieske} S.,  {Zeilinger} W.~W.,   {Chilingarian} I.,  2013, \mn@doi [\aap] {10.1051/0004-6361/201321466}, \href {https://ui.adsabs.harvard.edu/abs/2013A&A...557A..21S} {557, A21}

\bibitem[\protect\citeauthoryear{{Saulder} et~al.,}{{Saulder} et~al.}{2023}]{Saulder2023}
{Saulder} C.,  et~al., 2023, \mn@doi [\mnras] {10.1093/mnras/stad2200}, \href {https://ui.adsabs.harvard.edu/abs/2023MNRAS.tmp.2152S} {}

\bibitem[\protect\citeauthoryear{{Schlafly} et~al.,}{{Schlafly} et~al.}{2023}]{Schlafly2023}
{Schlafly} E.~F.,  et~al., 2023, \mn@doi [\aj] {10.3847/1538-3881/ad0832}, \href {https://ui.adsabs.harvard.edu/abs/2023AJ....166..259S} {166, 259}

\bibitem[\protect\citeauthoryear{{Scolnic}, {Boubel}, {Byrne}, {Riess}  \& {Anand}}{{Scolnic} et~al.}{2024}]{Scolnic2024}
{Scolnic} D.,  {Boubel} P.,  {Byrne} J.,  {Riess} A.~G.,   {Anand} G.~S.,  2024, \mn@doi [arXiv e-prints] {10.48550/arXiv.2412.08449}, \href {https://ui.adsabs.harvard.edu/abs/2024arXiv241208449S} {p. arXiv:2412.08449}

\bibitem[\protect\citeauthoryear{{Scolnic} et~al.,}{{Scolnic} et~al.}{2025}]{Scolnic2025}
{Scolnic} D.,  et~al., 2025, \mn@doi [\apjl] {10.3847/2041-8213/ada0bd}, \href {https://ui.adsabs.harvard.edu/abs/2025ApJ...979L...9S} {979, L9}

\bibitem[\protect\citeauthoryear{{Silber} et~al.,}{{Silber} et~al.}{2023}]{Silber2023}
{Silber} J.~H.,  et~al., 2023, \mn@doi [\aj] {10.3847/1538-3881/ac9ab1}, \href {https://ui.adsabs.harvard.edu/abs/2023AJ....165....9S} {165, 9}

\bibitem[\protect\citeauthoryear{{Springob} et~al.,}{{Springob} et~al.}{2014}]{Springob2014}
{Springob} C.~M.,  et~al., 2014, \mn@doi [\mnras] {10.1093/mnras/stu1743}, \href {https://ui.adsabs.harvard.edu/abs/2014MNRAS.445.2677S} {445, 2677}

\bibitem[\protect\citeauthoryear{{Taylor} et~al.,}{{Taylor} et~al.}{2023}]{Taylor2023}
{Taylor} E.~N.,  et~al., 2023, \mn@doi [The Messenger] {10.18727/0722-6691/5312}, \href {https://ui.adsabs.harvard.edu/abs/2023Msngr.190...46T} {190, 46}

\bibitem[\protect\citeauthoryear{{The Dark Energy Survey Collaboration}}{{The Dark Energy Survey Collaboration}}{2005}]{DES2005}
{The Dark Energy Survey Collaboration} 2005, \mn@doi [arXiv e-prints] {10.48550/arXiv.astro-ph/0510346}, \href {https://ui.adsabs.harvard.edu/abs/2005astro.ph.10346T} {pp astro--ph/0510346}

\bibitem[\protect\citeauthoryear{{Tully} \& {Fisher}}{{Tully} \& {Fisher}}{1977}]{Tully1977}
{Tully} R.~B.,  {Fisher} J.~R.,  1977, \aap, \href {http://adsabs.harvard.edu/abs/1977A%26A....54..661T} {54, 661}

\bibitem[\protect\citeauthoryear{{Tully}, {Courtois}  \& {Sorce}}{{Tully} et~al.}{2016}]{Tully2016}
{Tully} R.~B.,  {Courtois} H.~M.,   {Sorce} J.~G.,  2016, \mn@doi [\aj] {10.3847/0004-6256/152/2/50}, \href {http://adsabs.harvard.edu/abs/2016AJ....152...50T} {152, 50}

\bibitem[\protect\citeauthoryear{{Tully} et~al.,}{{Tully} et~al.}{2023}]{Tully2023}
{Tully} R.~B.,  et~al., 2023, \mn@doi [\apj] {10.3847/1538-4357/ac94d8}, \href {https://ui.adsabs.harvard.edu/abs/2023ApJ...944...94T} {944, 94}

\bibitem[\protect\citeauthoryear{{Turner}, {Blake}  \& {Ruggeri}}{{Turner} et~al.}{2023}]{Turner2023}
{Turner} R.~J.,  {Blake} C.,   {Ruggeri} R.,  2023, \mn@doi [\mnras] {10.1093/mnras/stac3256}, \href {https://ui.adsabs.harvard.edu/abs/2023MNRAS.518.2436T} {518, 2436}

\bibitem[\protect\citeauthoryear{{Valdes}, {Gupta}, {Rose}, {Singh}  \& {Bell}}{{Valdes} et~al.}{2004}]{Valdes2004}
{Valdes} F.,  {Gupta} R.,  {Rose} J.~A.,  {Singh} H.~P.,   {Bell} D.~J.,  2004, \mn@doi [\apjs] {10.1086/386343}, \href {https://ui.adsabs.harvard.edu/abs/2004ApJS..152..251V} {152, 251}

\bibitem[\protect\citeauthoryear{{Visser}}{{Visser}}{2004}]{Visser2004}
{Visser} M.,  2004, \mn@doi [Classical and Quantum Gravity] {10.1088/0264-9381/21/11/006}, \href {https://ui.adsabs.harvard.edu/abs/2004CQGra..21.2603V} {21, 2603}

\bibitem[\protect\citeauthoryear{{Vysotski{\v{i}}}, {Dolgov}  \& {Zel'Dovich}}{{Vysotski{\v{i}}} et~al.}{1977}]{Vysotski1977}
{Vysotski{\v{i}}} M.~I.,  {Dolgov} A.~D.,   {Zel'Dovich} Y.~B.,  1977, Soviet Journal of Experimental and Theoretical Physics Letters, \href {https://ui.adsabs.harvard.edu/abs/1977JETPL..26..188V} {26, 188}

\bibitem[\protect\citeauthoryear{{Watkins} et~al.,}{{Watkins} et~al.}{2023}]{Watkins2023}
{Watkins} R.,  et~al., 2023, \mn@doi [\mnras] {10.1093/mnras/stad1984}, \href {https://ui.adsabs.harvard.edu/abs/2023MNRAS.524.1885W} {524, 1885}

\bibitem[\protect\citeauthoryear{{Weinberg}}{{Weinberg}}{1972}]{Weinberg1972}
{Weinberg} S.,  1972, {Gravitation and Cosmology: Principles and Applications of the General Theory of Relativity}

\bibitem[\protect\citeauthoryear{{Weinberg}}{{Weinberg}}{2008}]{Weinberg2008}
{Weinberg} S.,  2008, {Cosmology}

\bibitem[\protect\citeauthoryear{{Whitford}, {Howlett}  \& {Davis}}{{Whitford} et~al.}{2023}]{Whitford2023}
{Whitford} A.~M.,  {Howlett} C.,   {Davis} T.~M.,  2023, \mn@doi [arXiv e-prints] {10.48550/arXiv.2306.11269}, \href {https://ui.adsabs.harvard.edu/abs/2023arXiv230611269W} {p. arXiv:2306.11269}

\bibitem[\protect\citeauthoryear{{Williams}, {Olszewski}, {Lesser}  \& {Burge}}{{Williams} et~al.}{2004}]{Williams2004}
{Williams} G.~G.,  {Olszewski} E.,  {Lesser} M.~P.,   {Burge} J.~H.,  2004, in {Moorwood} A. F.~M.,  {Iye} M.,  eds,  Society of Photo-Optical Instrumentation Engineers (SPIE) Conference Series Vol. 5492, Ground-based Instrumentation for Astronomy. pp 787--798, \mn@doi{10.1117/12.552189}

\bibitem[\protect\citeauthoryear{{Willmer}}{{Willmer}}{2018}]{Willmer2018}
{Willmer} C. N.~A.,  2018, \mn@doi [\apjs] {10.3847/1538-4365/aabfdf}, \href {https://ui.adsabs.harvard.edu/abs/2018ApJS..236...47W} {236, 47}

\bibitem[\protect\citeauthoryear{{Wright} et~al.,}{{Wright} et~al.}{2010}]{Wright2010}
{Wright} E.~L.,  et~al., 2010, \mn@doi [\aj] {10.1088/0004-6256/140/6/1868}, \href {https://ui.adsabs.harvard.edu/abs/2010AJ....140.1868W} {140, 1868}

\bibitem[\protect\citeauthoryear{{Zhou} et~al.,}{{Zhou} et~al.}{2021}]{Zhou2021}
{Zhou} R.,  et~al., 2021, \mn@doi [\mnras] {10.1093/mnras/staa3764}, \href {https://ui.adsabs.harvard.edu/abs/2021MNRAS.501.3309Z} {501, 3309}

\bibitem[\protect\citeauthoryear{{Zou} et~al.,}{{Zou} et~al.}{2017}]{Zou2017}
{Zou} H.,  et~al., 2017, \mn@doi [\pasp] {10.1088/1538-3873/aa65ba}, \href {https://ui.adsabs.harvard.edu/abs/2017PASP..129f4101Z} {129, 064101}

\makeatother
\end{thebibliography}






\bsp	
\label{lastpage}
\end{document}